\documentclass[12pt]{article}
\usepackage{amssymb}
\usepackage{rotating}

\renewcommand{\theequation}{\arabic{section}.\arabic{equation}}

\begin{document}



\def\a{\alpha}
\def\b{\beta}
\def\d{\delta}
\def\e{\epsilon}
\def\g{\gamma}
\def\h{\mathfrak{h}}
\def\k{\kappa}
\def\l{\lambda}
\def\o{\omega}
\def\p{\wp}
\def\r{\rho}
\def\t{\tau}
\def\s{\sigma}
\def\z{\zeta}
\def\x{\xi}
\def\V={{{\bf\rm{V}}}}
 \def\A{{\cal{A}}}
 \def\B{{\cal{B}}}
 \def\C{{\cal{C}}}
 \def\D{{\cal{D}}}
\def\G{\Gamma}
\def\K{{\cal{K}}}
\def\O{\Omega}
\def\R{\bar{R}}
\def\T{{\cal{T}}}
\def\L{\Lambda}
\def\f{E_{\tau,\eta}(sl_2)}
\def\E{E_{\tau,\eta}(sl_n)}
\def\Zb{\mathbb{Z}}
\def\Cb{\mathbb{C}}

\def\R{\overline{R}}

\def\beq{\begin{equation}}
\def\eeq{\end{equation}}
\def\bea{\begin{eqnarray}}
\def\eea{\end{eqnarray}}
\def\ba{\begin{array}}
\def\ea{\end{array}}
\def\no{\nonumber}
\def\le{\langle}
\def\re{\rangle}
\def\lt{\left}
\def\rt{\right}

\newtheorem{Theorem}{Theorem}
\newtheorem{Definition}{Definition}
\newtheorem{Proposition}{Proposition}
\newtheorem{Lemma}{Lemma}
\newtheorem{Corollary}{Corollary}
\newcommand{\proof}[1]{{\bf Proof. }
        #1\begin{flushright}$\Box$\end{flushright}}

\baselineskip=20pt

\newfont{\elevenmib}{cmmib10 scaled\magstep1}
\newcommand{\preprint}{
   \begin{flushleft}
   \end{flushleft}\vspace{-1.3cm}
   \begin{flushright}\normalsize
   \end{flushright}}
\newcommand{\Title}[1]{{\baselineskip=26pt
   \begin{center} \Large \bf #1 \\ \ \\ \end{center}}}

\newcommand{\Author}{\begin{center}
   \large \bf
Xiaotian Xu${}^{a}$, Junpeng Cao${}^{a,b,c,d}$,  Yi Qiao${}^{a,e}$, Wen-Li
Yang${}^{d,e,f,g}\footnote{Corresponding author: wlyang@nwu.edu.cn}$, Kangjie Shi${}^e$ and Yupeng
Wang${}^{a,d,h}\footnote{Corresponding author: yupeng@iphy.ac.cn}$
 \end{center}}

\newcommand{\Address}{\begin{center}

     ${}^a$ Beijing National Laboratory for Condensed Matter
           Physics, Institute of Physics, Chinese Academy of Sciences, Beijing
           100190, China\\
     ${}^b$ Songshan Lake Materials Laboratory, Dongguan, Guangdong 523808, China \\
     ${}^c$ School of Physical Sciences, University of Chinese Academy of
Sciences, Beijing, China\\
     ${}^d$ Peng Huanwu Center for Fundamental Theory, Xian 710127, China\\
     ${}^e$ Institute of Modern Physics, Northwest University,
     Xian 710127, China\\
     ${}^f$ Shaanxi Key Laboratory for Theoretical Physics Frontiers,  Xian 710127, China\\
     ${}^g$ Physics school, Northwest University,  Xian 710127, China\\
     ${}^h$ The Yangtze River Delta Physics Research Center, Liyang, Jiangsu, China
\end{center}}

\newcommand{\Accepted}[1]{\begin{center}
   {\large \sf #1}\\ \vspace{1mm}{\small \sf Accepted for Publication}
   \end{center}}

\preprint
\thispagestyle{empty}
\bigskip\bigskip\bigskip

\Title{Graded off-diagonal Bethe ansatz solution of the $SU(2|2)$  spin chain model with generic integrable boundaries} \Author

\Address
\vspace{1cm}

\begin{abstract}
\bigskip
The graded off-diagonal Bethe ansatz method is proposed to study  supersymmetric quantum integrable models (i.e., quantum integrable models associated with superalgebras). As an example,
the exact solutions of the $SU(2|2)$ vertex model with both periodic and generic open boundary conditions are constructed.
By generalizing the fusion techniques to the supersymmetric case, a closed set of operator product identities about the transfer matrices are derived, which allows us to give the eigenvalues in terms of
homogeneous or inhomogeneous $T-Q$ relations. The method and results provided in this paper can be generalized to other high rank supersymmetric quantum integrable models.

\vspace{1truecm} \noindent {\it PACS:} 75.10.Pq, 02.30.Ik, 71.10.Pm

\noindent {\it Keywords}: Bethe Ansatz; Lattice Integrable Models; $T-Q$ Relation
\end{abstract}
\newpage

\section{Introduction}
\label{intro} \setcounter{equation}{0}

Quantum integrable models \cite{Bax82} play important roles in fields of theoretical physics, condensed matter physics, field theory and
mathematical physics, since exact solutions of those models may provide useful benchmarks to understand a variety of many-body problems.
During the past several decades, much attention has been paid to obtain exact solutions of integrable systems with unusual boundary conditions.
With the development of topological physics and string theory, study on off-diagonal boundaries becomes an interesting issue.
Many interesting phenomena such as edge states, Majorana zero modes, and topological excitations have been found.

Due to the existence of off-diagonal elements contained in boundaries, particle numbers with different intrinsic degrees of freedom are
not conserved anymore and the usual $U(1)$ symmetry is broken. This leads to absence of a proper reference state which is crucial in the conventional Bethe ansatz scheme. To overcome this problem,
several interesting methods \cite{Alc87,Skl88, Nep02, Cao03, Yan04, Gie05, Yan07, Bas06, Bas07, Bas10, Bas13,Fra08, Fra11, Nic13,cao13, wang15, Bel13, Bel15, Pim15, Ava15} are proposed. A remarkable one is the off-diagonal Bethe ansatz (ODBA) \cite{cao13, wang15}, which allow us to construct the exact spectrum systematically. The nested ODBA  has also been developed to deal with the models with different Lie algebras such as $A_n$ \cite{Cao14, Cao15}, $A_2^{(2)}$ \cite{Hao14}, $B_2$ \cite{ Li_119}, $C_2$ \cite{Li_219} and $D_3^{(1)}$ \cite{Li_319}.
Nevertheless, there exists another kind of high rank integrable models which are related to superalgebras \cite{Fra00} such as the
$SU(m|n)$ model, the Hubbard model, and the supersymmetric $t-J$ model.
The $SU(m|n)$ model has many applications in AdS/CFT correspondence \cite{Mal99,Bei12}, while the Hubbard and $t-J$ model have many applications in the strongly correlated
electronic theory. These models with $U(1)$ symmetry have been studied extensively \cite{yb, Ess05,Perk81,Vega91,Yue96,Yue1,Yue2,Yue3,Yue4,Yue5,Mar97}.
A general method to approach
such kind of  models with off-diagonal boundaries is  still missing.

In this paper, we develop a graded version of nested ODBA to study supersymmetric integrable models (integrable models associated with superalgebras). As an example, the $SU(2|2)$ model with both periodic and off-diagonal boundaries is studied.
The structure of the paper is as follows. In section 2, we study the $SU(2|2)$ model with periodic boundary condition.
A closed set of  operator identities is constructed by using the fusion procedure. These identities allow us to characterize the eigenvalues of the transfer matrices in terms of
homogeneous $T-Q$ relation. In section 3,
we study the model with generic open boundary conditions. It is demonstrated that similar identities can be constructed and the spectrum can be expressed in terms of inhomogeneous $T-Q$ relation. Section 4 is attributed to concluding remarks. Some technical details can be found in the appendices.

\section{$SU(2|2)$ model with periodic boundary condition}
\label{c2} \setcounter{equation}{0}

\subsection{The system}

Let ${V}$ denote a $4$-dimensional graded linear space with a basis $\{|i\rangle|i=1,\cdots,4\}$, where the Grassmann parities
are $p(1)=0$, $p(2)=0$, $p(3)=1$ and $p(4)=1$, which endows the fundamental
representation of the $SU(2|2)$ Lie superalgebra. The dual space is spanned by the dual basis $\{\langle i|\,\,|i=1,\cdots,4\}$ with an inner product: $\langle i|j\rangle=\delta_{ij}$.
Let us further introduce the ${Z_{2}}$-graded $N$-tensor space ${V}\otimes { V}\otimes\cdots{V}$ which has a basis $\{|i_1,i_2,\cdots,i_N\rangle=|i_N\rangle_N\,\cdots|i_2\rangle_2\,|i_1\rangle_1\,|\,i_l=1,\cdots,4;\,l=1,\cdots,N\}$,
and its dual with a basis  $\{\langle i_1,i_2,\cdots,i_N|=\langle i_1|_1\,\langle i_2|_2\,\cdots\langle i_N|_N\,|\,i_l=1,\cdots,4;\,l=1,\cdots,N\}$.

For the matrix $A_j\in {\rm End}({ V_j})$, $A_j$ is a super
embedding operator in the ${Z_{2}}$-graded $N$-tensor space ${V}\otimes { V}\otimes\cdots{V}$, which acts as $A$ on the $j$-th
space and as identity on the other factor spaces. For the matrix $R_{ij}\in {\rm
End}({ V_i}\otimes { V_j})$, $R_{ij}$ is a super embedding
operator in the ${Z_{2}}$ graded tensor space, which acts as
identity on the factor spaces except for the $i$-th and $j$-th ones.
The super tensor product of two operators
is the graded one satisfying the rule\footnote{For $A=\sum_{\alpha,\,\beta}A_{\beta}^{\alpha}{|\beta\rangle}{\langle\alpha|}$ and $B=\sum_{\delta,\,\gamma}B_{\delta}^{\gamma}{|\delta\rangle}{\langle\gamma|}$, the super tensor product $A\otimes B=\sum_{\a,\b,\gamma,\delta}(A_{\beta}^{\alpha}{|\beta\rangle}_1{\langle\alpha|}_{1})\,\, (B_{\delta}^{\gamma}{|\delta\rangle}_2{\langle\gamma|}_2)=\sum_{\a,\b,\gamma,\delta}(-1)^{p(\delta)[p(\alpha)+p(\beta)]}A_{\beta}^{\alpha}B_{\delta}^{\gamma}{|\delta\rangle}_2
{|\beta\rangle}_1\,{\langle\alpha|}_{1}{\langle\gamma|}_2$.} $(A\otimes B)_{\beta \delta}^{\alpha
\gamma}=(-1)^{[p(\alpha)+p(\beta)]p(\delta)}A^{\alpha}_{\beta}B^{\gamma}_{\delta}$
\cite{Gra13}.

The supersymmetric $SU(2|2)$ model is described by the $16\times 16$ $R$-matrix
\begin{equation}
\label{rm}
R_{12}(u)
=\left(
                \begin{array}{cccc|cccc|cccc|cccc}
      u+\eta &  &  &  &  &  &  &  &  &  &  &  &  &  &  &  \\
      & u &  &  & \eta &  &  &  &  &  &  &  &  &  &  &  \\
      &  & u &  &  &  &  &  & \eta &  &  &  &  &  &  &  \\
      &  &  & u &  &  &  &  &  &  &  &  & \eta &  &  &  \\
     \hline
      & \eta &  &  & u &  &  &  &  &  &  &  &  &  &  &  \\
      &  &  &  &  & u+\eta &  &  &  &  &  &  &  &  &  &  \\
      &  &  &  &  &  & u &  &  & \eta &  &  &  &  &  &  \\
      &  &  &  &  &  &  & u &  &  &  &  &  & \eta &  &  \\
     \hline
      &  & \eta &  &  &  &  &  & u &  &  &  &  &  &  &  \\
      &  &  &  &  &  & \eta &  &  & u &  &  &  &  &  &  \\
      &  &  &  &  &  &  &  &  &  & u-\eta &  &  &  &  &  \\
      &  &  &  &  &  &  &  &  &  &  & u &  &  & -\eta &  \\
    \hline
      &  &  & \eta &  &  &  &  &  &  &  &  & u &  &  &  \\
      &  &  &  &  &  &  & \eta &  &  &  &  &  & u &  &  \\
      &  &  &  &  &  &  &  &  &  &  & -\eta &  &  & u &  \\
      &  &  &  &  &  &  &  &  &  &  &  &  &  &  & u-\eta \\
   \end{array}
 \right),
\end{equation}
where $u$ is the spectral parameter and $\eta$ is the crossing parameter.
The $R$-matrix (\ref{rm}) enjoys the following properties
\begin{eqnarray}
{\rm regularity}&:&R _{12}(0)=\eta P_{12},\nonumber\\[4pt]
{\rm unitarity}&:&R_{12}(u)R_{21}(-u) = \rho_1(u)\times {\rm id},\nonumber\\[4pt]
{\rm crossing-unitarity}&:&R_{12}^{st_1}(-u)R_{21}^{st_1}(u)=\rho_2(u)\times {\rm id},\nonumber
\end{eqnarray}
where $P_{12}$ is the $Z_2$-graded permutation operator with the definition
\begin{eqnarray}
P_{\beta_{1}\beta_{2}}^{\alpha_{1}\alpha_{2}}=(-1)^{p(\alpha_{1})p(\alpha_{2})} \delta_{\alpha_{1}\beta_{2}}
\delta_{\beta_{1}\alpha_{2}},
\end{eqnarray}
$R_{21}(u)=P_{12}R_{12}(u)P_{12}$,
$st_i$ denotes the super transposition in the $i$-th space
$(A^{st_i})_{ij}=A_{ji}(-1)^{p(i)[p(i)+p(j)]}$, and the functions $\rho_1(u)$ and
$\rho_2(u)$ are given by
\begin{eqnarray}
\rho_1(u)=-({u}-\eta)({u}+\eta), \quad
\rho_2(u)=-u^2.
\end{eqnarray}
The $R$-matrix (\ref{rm}) satisfies the graded Yang-Baxter equation (GYBE) \cite{Kul1, Kul86}
\begin{eqnarray}
R_{12}(u-v)R_{13}(u)R_{23}(v)=R_{23}(v)R_{13}(u)R_{12}(u-v)\label{YBE}.
\end{eqnarray}
In terms of the matrix entries, GYBE (\ref{YBE}) reads
\bea
&&\sum_{\beta_1,\beta_2,\beta_3}R(u-v)_{\beta_1\beta_2}^{\alpha_1\alpha_2}R(u)_{\gamma_1\beta_3}^{\beta_1\alpha_3}
R(v)_{\gamma_2\gamma_3}^{\beta_2\beta_3}(-1)^{(p(\beta_1)+p(\gamma_1))p(\beta_2)}\no\\[4pt]
&&=\sum_{\beta_1,\beta_2,\beta_3}R(v)_{\beta_2\beta_3}^{\alpha_2\alpha_3}R(u)_{\beta_1\gamma_3}^{\alpha_1\beta_3}
R(u-v)_{\gamma_1\gamma_2}^{\beta_1\beta_2}(-1)^{(p(\alpha_1)+p(\beta_1))p(\beta_2)}.
\eea
For the periodic boundary condition, we introduce the ``row-to-row"  (or one-row) monodromy matrix
$T_0(u)$
\begin{eqnarray}
T_0 (u)=R _{01}(u-\theta_1)R _{02}(u-\theta_2)\cdots R _{0N}(u-\theta_N),\label{T1}
\end{eqnarray}
where the subscript $0$ means the auxiliary space $V_0$,
the other tensor space $V^{\otimes N}$ is the physical or quantum space, $N$ is the number of sites and $\{\theta_j|j=1,\cdots,N\}$ are the inhomogeneous parameters.
In the auxiliary space, the monodromy matrix (\ref{T1}) can be written as a $4\times 4$ matrix with operator-valued elements
acting on ${\rm V}^{\otimes N}$.
The explicit forms of the elements of monodromy matrix (\ref{T1}) are
\bea
\Big\{[T_0(u)]^{a}_b\Big\}_{\beta_1\cdots\beta_N}^{\alpha_1\cdots\alpha_N}&=&\sum_{c_2,\cdots,c_N}R_{0N}(u)_{c_N\beta_N}^{a\alpha_N}\cdots R_{0j}(u)_{c_j\beta_j}^{c_{j+1}\alpha_j}\cdots R_{01}(u)_{b\beta_1}^{c_2\alpha_1}\no\\[4pt]
&&\times (-1)^{\sum_{j=2}^{N}(p(\alpha_j)+p(\beta_j))\sum_{i=1}^{j-1}p(\alpha_i)}.
\eea
The monodromy matrix $T_0(u)$ satisfies the graded Yang-Baxter relation
\begin{eqnarray}
R _{12}(u-v)T_1 (u) T_2 (v) =T_2 (v)T_1 (u)R_{12}(u-v).
\label{ybe}
\end{eqnarray}
The transfer matrix $t_p(u)$ of the system is defined as the super partial trace of the monodromy matrix in the auxiliary space
\begin{eqnarray}
t_p(u)=str_0\{T_0 (u)\}=\sum_{\alpha=1}^{4}(-1)^{p(\alpha)}[T_0(u)]_{\alpha}^{\alpha}.
\end{eqnarray}
From the graded Yang-Baxter relation (\ref{ybe}), one can prove that the transfer matrices with different spectral parameters
commute with each other, $[t_p(u), t_p(v)]=0$. Thus $t_p(u)$ serves as the
generating functional of all the conserved quantities, which ensures the
integrability of the system. The model Hamiltonian is constructed by \cite{Yue1}
\bea
H_p=\frac{\partial \ln t_p(u)}{\partial u}|_{u=0,\{\theta_j\}=0}.\label{peri-Ham}
\eea

\subsection{Fusion}

One of the wonderful properties of $R$-matrix is that it may degenerate to the projection
operators at some special points, which makes it possible to do the fusion procedure
\cite{Kul81, Kul82, Kar79, Kir86, Kir87, Tsu97}. It is easy to check that the $R$-matrix (\ref{rm}) has two degenerate points.
The first one is $u=\eta$. At which, we have
\bea
R _{12}(\eta)= 2\eta P_{12}^{(8)},\label{Int-R1}\eea
where $P_{12}^{(8)}$ is a 8-dimensional supersymmetric projector
\bea
P_{12}^{(8)}=\sum_{i=1}^{8}|f_i\rangle  \langle f_i|, \label{1-project}\eea
and the corresponding basis vectors are
\bea
&&|f_1\rangle= |11\rangle, \quad |f_2\rangle =\frac{1}{\sqrt{2}}(|12\rangle +|21\rangle ), \quad|f_3\rangle =|22\rangle,\nonumber\\
&&|f_4\rangle=\frac{1}{\sqrt{2}}(|34\rangle -|43\rangle ),\quad |f_5\rangle=\frac{1}{\sqrt{2}}(|13\rangle +|31\rangle ),\quad|f_6\rangle=\frac{1}{\sqrt{2}}(|14\rangle +|41\rangle ),\nonumber\\
&&|f_7\rangle=\frac{1}{\sqrt{2}}(|23\rangle +|32\rangle ),\quad|f_8\rangle= \frac{1}{\sqrt{2}}(|24\rangle +|42\rangle ),\no
\eea
with the corresponding parities
\bea
p(f_1)=p(f_2)=p(f_3)=p(f_4)=0, \quad p(f_5)=p(f_6)=p(f_7)=p(f_8)=1. \no
\eea
The operator $P_{12}^{(8)}$ projects the original 16-dimensional tensor space $V_1\otimes V_2$ into
a new 8-dimensional projected space
spanned by $\{|f_i\rangle|i=1,\cdots,8\}$.
Taking the fusion by the operator (\ref{1-project}), we construct the fused $R$-matrices
\bea
&&R_{\langle 12\rangle 3}(u)=(u+\frac{1}{2}\eta)^{-1}P^{ (8) }_{12}R _{23}(u-\frac{1}{2}\eta)R _{13}(u+\frac{1}{2}\eta)P^{ (8) }_{12}\equiv R_{\bar 1 3}(u), \label{hhgg-1}\\[4pt]
&&R_{3 \langle 21\rangle}(u)=(u+\frac{1}{2}\eta)^{-1}P^{ (8) }_{21}R _{32}(u-\frac{1}{2}\eta)R _{31}(u+\frac{1}{2}\eta)P^{ (8) }_{21}\equiv R_{3\bar 1}(u), \label{hhgg-2}
\eea
where $P^{ (8) }_{21}$ can be obtained from $P^{ (8) }_{12}$ by exchanging  $V_1$ and $V_2$. For simplicity, we denote the projected space as
$V_{\bar 1}=V_{\langle12\rangle}=V_{\langle21\rangle}$.
The fused $R$-matrix ${R}_{\bar{1}2}(u)$ is a $32\times 32$ matrix defined in the tensor space $V_{\bar 1}\otimes V_2$ and has the properties
\bea
&&R_{\bar{1}2}(u)R_{2\bar{1}}(-u)=\rho_3(u)\times {\rm id}, \no\\[4pt]
&&R_{\bar{1}2}(u)^{st_{\bar 1}} R_{2\bar{1}}(-u)^{st_{\bar 1}}=\rho_4(u)\times {\rm id},
\eea
where
\bea
\rho_3(u)=-(u+\frac{3}{2}\eta)(u-\frac{3}{2}\eta),\quad \rho_4(u)=-(u+\frac{1}{2}\eta)(u-\frac{1}{2}\eta).
\eea
From GYBE (\ref{YBE}), one can prove that the following fused graded Yang-Baxter equations hold
\bea
R_{\bar{1}2}(u-v) R_{\bar{1}3}(u) R_{23}(v)
=R_{23}(v)R_{\bar{1}3}(u)R_{\bar{1}2}(u-v).\label{fuse-qybe1}
\eea
It is easy to check that the elements of fused $R$-matrices $R_{\bar{1}2}(u)$ and $R_{2\bar{1}}(u)$
are degree one polynomials of $u$.

At the point of $u=-\frac{3}{2}\eta$, the fused $R$-matrix $R_{\bar{1}2}(u)$ can also be written as a projector
\bea R_{\bar{1}2}(-\frac{3}{2}\eta)= -3\eta P^{(20) }_{{\bar 1}2},\label{Fusion-5-4}
\eea
where $P^{(20) }_{\bar{1} 2}$ is a 20-dimensional supersymmetric projector
\bea
P^{(20) }_{\bar{1}2}=\sum_{i=1}^{20} |\phi_i\rangle  \langle \phi_i|, \label{cjjc}\eea
with the basis vectors
\bea
&&|\phi_1\rangle =\frac{1}{\sqrt{3}}(\sqrt{2}|f_1\rangle\otimes|2\rangle -|f_2\rangle\otimes|1\rangle),\quad |\phi_2\rangle=\frac{1}{\sqrt{3}}( |f_2\rangle\otimes|2\rangle -\sqrt{2}|f_3\rangle\otimes|1\rangle),\nonumber\\[4pt]
&&|\phi_{3}\rangle=\frac{1}{\sqrt{6}}(2|f_6\rangle\otimes|3\rangle+|f_5\rangle\otimes|4\rangle +|f_4\rangle\otimes|1\rangle ),\quad|\phi_{4}\rangle=\frac{1}{\sqrt{2}}(|f_5\rangle\otimes|4\rangle -|f_4\rangle\otimes|1\rangle ),\nonumber\\[4pt]
&&|\phi_{5}\rangle =\frac{1}{\sqrt{6}}(|f_8\rangle\otimes|3\rangle+2|f_4\rangle\otimes|2\rangle -|f_7\rangle\otimes|4\rangle ),\quad |\phi_{6}\rangle=\frac{1}{\sqrt{2}}(|f_7\rangle\otimes|4\rangle +|f_8\rangle\otimes|3\rangle ),\nonumber\\[4pt]
&&|\phi_{7}\rangle =|f_5\rangle\otimes|3\rangle ,\quad |\phi_{8}\rangle=|f_7\rangle\otimes|3\rangle,\quad |\phi_{9}\rangle=|f_6\rangle\otimes|4\rangle ,\quad|\phi_{10}\rangle =|f_8\rangle\otimes|4\rangle,\nonumber\\[4pt]
&&|\phi_{11}\rangle=\frac{1}{\sqrt{3}}(\sqrt{2}|f_1\rangle\otimes|3\rangle -|f_5\rangle\otimes|1\rangle ),\quad |\phi_{12}\rangle =\frac{1}{\sqrt{3}}( \sqrt{2}|f_1\rangle\otimes|4\rangle -|f_6\rangle\otimes|1\rangle),\nonumber\\[4pt]
&&|\phi_{13}\rangle =\frac{1}{\sqrt{6}}(|f_7\rangle\otimes|1\rangle+|f_2\rangle\otimes|3\rangle -2|f_5\rangle\otimes|2\rangle ),\quad |\phi_{14}\rangle=\frac{1}{\sqrt{2}}(|f_2\rangle\otimes|3\rangle -|f_7\rangle\otimes|1\rangle )\nonumber\\[4pt]
&&|\phi_{15}\rangle =\frac{1}{\sqrt{6}}(|f_8\rangle\otimes|1\rangle+|f_2\rangle\otimes|4\rangle -2|f_6\rangle\otimes|2\rangle ),\quad |\phi_{16}\rangle=\frac{1}{\sqrt{2}}(|f_2\rangle\otimes|4\rangle -|f_8\rangle\otimes|1\rangle ),\nonumber\\[4pt]
&&|\phi_{17}\rangle =\frac{1}{\sqrt{3}}(\sqrt{2}|f_3\rangle\otimes|3\rangle -|f_7\rangle\otimes|2\rangle ),\quad|\phi_{18}\rangle=\frac{1}{\sqrt{3}}(\sqrt{2}|f_3\rangle\otimes|4\rangle -|f_8\rangle\otimes|2\rangle ),\nonumber\\[4pt]
&&|\phi_{19}\rangle =|f_4\rangle\otimes|3\rangle ,\quad |\phi_{20}\rangle=|f_4\rangle\otimes|4\rangle.\no
\eea
The corresponding parities of the basis vectors are
\bea
p(\phi_1)=p(\phi_2)=\cdots =p(\phi_{10})=0,\quad p(\phi_{11})=p(\phi_{12})=\cdots =p(\phi_{20})=1. \no
\eea
The operator $P^{(20)}_{{\bar 1}2}$ is a projector on the 32-dimensional product space $V_{\bar 1}\otimes V_2$ which projects $V_{\bar 1}\otimes V_2$ into
its 20-dimensional subspace
spanned by $\{|\phi_i\rangle, i=1,\cdots,20\}$.

Taking the fusion by the projector $P^{(20)}_{{\bar 1}2}$, we obtain another new fused $R$-matrix
\bea &&
{R}_{\langle {\bar 1}2\rangle 3}(u)
=(u-\eta)^{-1}P^{(20) }_{2\bar{1}}R_{\bar{1}3}(u+\frac{1}{2}\eta) R_{23}(u-\eta)P^{(20)}_{2\bar{1} }\equiv {R}_{\tilde 1 3}(u), \label{fu-2}\\[4pt]
&&{R}_{3\langle 2{\bar 1}\rangle}(u)
=(u-\eta)^{-1}P^{(20) }_{{\bar 1}2}R_{3\bar{1}}(u+\frac{1}{2}\eta) R_{32}(u-\eta)P^{(20)}_{{\bar 1}2}\equiv {R}_{3\tilde 1}(u), \label{fu-22}
\eea
where $P^{(20)}_{2{\bar 1}}$ can be obtained from $P^{(20)}_{{\bar 1}2}$ by exchanging  $V_{\bar 1}$ and $V_2$.
For simplicity, we denote the projected subspace as $V_{\tilde 1}=V_{\langle\bar 12\rangle}=V_{\langle2\bar 1\rangle}$.
The fused $R$-matrix $R_{\tilde{1}2}(u)$ is a $80\times 80$ matrix defined in the tensor space $V_{\tilde 1}\otimes V_2$ and
satisfies following graded Yang-Baxter equations
\bea
R_{{\tilde 1}2}(u-v) R_{{\tilde 1}3}(u) R_{{2}3}(v)= R_{{2}3}(v) R_{{\tilde 1}3}(u)R_{{\tilde 1}2}(u-v).\label{sdfu-22}
\eea
The elements of fused $R$-matrix $R_{\tilde{1} 2}(u)$ are also degree one polynomials of $u$.

The second degenerate point of $R$-matrix (\ref{rm}) is $u=-\eta$. At which we have
\bea
 R_{12}(-\eta)= -2\eta \bar P_{12}^{(8)}=-2\eta(1-P_{12}^{(8)}),\label{2-pewrerroject}
\eea
where $\bar P_{12}^{(8)}$ is an 8-dimensional supersymmetric projector in terms of
\bea
\bar P_{12}^{(8)}=\sum_{i=1}^{8}|g_i\rangle  \langle g_i|, \label{2-project}\eea
with
\bea
&&|g_1\rangle= \frac{1}{\sqrt{2}}(|12\rangle -|21\rangle ), \quad |g_2\rangle=|33\rangle, \quad |g_3\rangle= \frac{1}{\sqrt{2}}(|34\rangle +|43\rangle ),\nonumber\\
&&|g_4\rangle=|44\rangle,\quad |g_5\rangle =\frac{1}{\sqrt{2}}(|13\rangle -|31\rangle ), \quad |g_6\rangle=\frac{1}{\sqrt{2}}(|14\rangle -|41\rangle ) \nonumber\\
&&|g_7\rangle=\frac{1}{\sqrt{2}}(|23\rangle -|32\rangle ),\quad |g_8\rangle =\frac{1}{\sqrt{2}}(|24\rangle -|42\rangle ). \label{fuse-q1ybe2}
\eea
The corresponding parities are
\bea
p(g_1)=p(g_2)=p(g_3)=p(g_4)=0,\quad p(g_5)=p(g_6)=p(g_7)=p(g_8)=1. \no
\eea
The operator $\bar P_{12}^{(8)}$ projects the 16-dimensional product space $V_1\otimes V_2$ into
a new 8-dimensional projected space spanned by $\{|g_i\rangle|i=1,\cdots,8\}$.

Taking the fusion by the projector $\bar P_{12}^{(8)}$, we obtain the fused $R$-matrices
\bea
&&R_{\langle 12\rangle^\prime 3}(u)=(u-\frac{1}{2}\eta)^{-1}\bar P^{ (8) }_{12}R _{23}(u+\frac{1}{2}\eta)R _{13}(u-\frac{1}{2}\eta)\bar P^{ (8) }_{12}\equiv R_{\bar{1}^\prime 3}(u), \label{hhgg-3}\\[4pt]
&&R_{3 \langle 21\rangle^\prime}(u)=(u-\frac{1}{2}\eta)^{-1}\bar P^{ (8) }_{21}R _{32}(u+\frac{1}{2}\eta)R _{31}(u-\frac{1}{2}\eta)\bar P^{ (8) }_{21}\equiv R_{3\bar{1}^\prime }(u).\label{hhgg-4}
\eea
For simplicity, we denote the projected space as
$V_{\bar 1^\prime}=V_{\langle12\rangle^\prime}=V_{\langle21\rangle^\prime}$.
The fused $R$-matrix $R_{\bar{1}^\prime 2}(u)$ is a $32\times 32$ matrix defined in the product space $V_{\bar{1}^\prime }\otimes V_2$ and possesses the properties
 \bea
&&R_{\bar{1}^\prime 2}(u) R_{2\bar{1}^\prime }(-u)=\rho_5(u)\times {\rm id}, \no\\[4pt]
&&R_{\bar{1}^\prime 2}(u)^{st_{{\bar 1}^\prime }} R_{2\bar{1}^\prime }(-u)^{st_{{\bar 1}^\prime }}=\rho_6(u)\times {\rm id}, \no\\[6pt]
&&R_{\bar{1}^\prime 2}(u-v) R_{\bar{1}^\prime 3}(u) R_{23}(v)
=R_{23}(v)R_{\bar{1}^\prime 3}(u)R_{\bar{1}^\prime 2}(u-v),\label{fuse-qybe2}
\eea
where
\bea
\rho_5(u)=-(u-\frac{3}{2}\eta)(u+\frac{3}{2}\eta),\quad \rho_6(u)=-(u-\frac{1}{2}\eta)(u+\frac{1}{2}\eta).
\eea

Now, we consider the fusions of $R_{\bar{1}^\prime 2}(u)$, which include two different cases. One is the fusion in the auxiliary space $V_{\bar 1}$ and
the other is the fusion in the quantum space $V_2$. Both are necessary to close the fusion processes.

We first introduce the fusion in the auxiliary space. At the point $u=\frac{3}{2}\eta$, we have
\bea R_{\bar{1}^\prime 2}(\frac{3}{2}\eta)= 3\eta P^{(20) }_{\bar{1}^\prime 2},\label{Fusion-20-1}\eea
where $P^{(20) }_{\bar{1}^\prime 2}$ is a 20-dimensional supersymmetric projector with the form of
\bea
P^{(20) }_{\bar{1}^\prime 2}=\sum_{i=1}^{20} |\tilde{\phi}_i\rangle  \langle \tilde{\phi}_i|, \label{xiao1}\eea
and the corresponding vectors are
\bea
&&|\tilde{\phi}_1\rangle =|g_1\rangle\otimes|1\rangle,\quad |\tilde{\phi}_2\rangle=|g_1\rangle\otimes|2\rangle,\nonumber\\[4pt]
&&|\tilde{\phi}_3\rangle=\frac{1}{\sqrt{2}}(|g_3\rangle\otimes|1\rangle -|g_5\rangle\otimes|4\rangle ),\quad |\tilde{\phi}_{4}\rangle=\frac{1}{\sqrt{6}}( |g_5\rangle\otimes4\rangle+|g_3\rangle\otimes|1\rangle-2|g_6\rangle\otimes|3\rangle ),\nonumber\\[4pt]
&&|\tilde{\phi}_{5}\rangle=\frac{1}{\sqrt{2}}(|g_8\rangle\otimes|3\rangle -|g_7\rangle\otimes|4\rangle ) ,\quad |\tilde{\phi}_{6}\rangle=\frac{1}{\sqrt{6}}(2|g_3\rangle\otimes|2\rangle-|g_7\rangle\otimes4\rangle -|g_8\rangle\otimes|3\rangle ),\nonumber\\[4pt]
&&|\tilde{\phi}_{7}\rangle =\frac{1}{\sqrt{3}}(\sqrt{2}|g_2\rangle\otimes|1\rangle -|g_5\rangle\otimes|3\rangle ),\quad |\tilde{\phi}_{8}\rangle=\frac{1}{\sqrt{3}}(\sqrt{2}|g_2\rangle\otimes|2\rangle -|g_7\rangle\otimes|3\rangle ),\nonumber\\[4pt]
&&|\tilde{\phi}_{9}\rangle=\frac{1}{\sqrt{3}}(\sqrt{2}|g_4\rangle\otimes|1\rangle -|g_6\rangle\otimes|4\rangle ),\quad|\tilde{\phi}_{10}\rangle =\frac{1}{\sqrt{3}}(\sqrt{2}|g_4\rangle\otimes|2\rangle -|g_8\rangle\otimes|4\rangle),\nonumber\\[4pt]
&&|\tilde{\phi}_{11}\rangle=|g_5\rangle\otimes|1\rangle,\quad|\tilde{\phi}_{12}\rangle =|g_6\rangle\otimes|1\rangle),\nonumber\\[4pt]
&&|\tilde{\phi}_{13}\rangle =\frac{1}{\sqrt{2}}(|g_7\rangle\otimes|1\rangle-|g_1\rangle\otimes|3\rangle ),\quad |\tilde{\phi}_{14}\rangle=\frac{1}{\sqrt{6}}(|g_7\rangle\otimes|1\rangle+2|g_5\rangle\otimes2\rangle +|g_1\rangle\otimes|3\rangle )\nonumber\\[4pt]
&&|\tilde{\phi}_{15}\rangle=\frac{1}{\sqrt{2}}(|g_8\rangle\otimes|1\rangle -|g_1\rangle\otimes|4\rangle ),\quad |\tilde{\phi}_{16}\rangle=\frac{1}{\sqrt{6}}(|g_6\rangle\otimes|2\rangle +2|g_8\rangle\otimes1\rangle +|g_1\rangle\otimes|4\rangle ),\nonumber\\[4pt]
&&|\tilde{\phi}_{17}\rangle=|g_7\rangle\otimes|2\rangle,\quad |\tilde{\phi}_{18}\rangle=|g_8\rangle\otimes|2\rangle,\nonumber\\[4pt]
&&|\tilde{\phi}_{19}\rangle=\frac{1}{\sqrt{3}}(|g_3\rangle\otimes|3\rangle-\sqrt{2}|g_2\rangle\otimes|4\rangle ),\quad|\tilde{\phi}_{20}\rangle=\frac{1}{\sqrt{3}}(\sqrt{2}|g_4\rangle\otimes|3\rangle -|g_3\rangle\otimes|4\rangle).\no
\eea
The parities read
\bea
p(\tilde\phi_1)=p(\tilde\phi_2)=\cdots=p(\tilde\phi_{10})=0,\quad p(\tilde\phi_{11})=p(\tilde\phi_{12})=\cdots=p(\tilde\phi_{20})=1.\no
\eea
The operator $P^{(20)}_{{\bar 1}^\prime 2}$ projects the 32-dimensional product space $V_{{\bar 1}^\prime} \otimes V_2$ into
a 20-dimensional projected space spanned by $\{|\tilde \phi_i\rangle, i=1,\cdots,20\}$.
Taking the fusion by the projector $P^{(20)}_{{\bar 1}^\prime 2}$, we obtain the following fused $R$-matrices
\bea
&&{R}_{\langle {\bar 1}^\prime 2\rangle 3}(u)=
(u+\eta)^{-1}P^{(20) }_{2\bar{1}^\prime }R_{\bar{1}^\prime 3}(u-\frac{1}{2}\eta) R_{23}(u+\eta)P^{(20)}_{2\bar{1}^\prime }\equiv R_{{\tilde 1}^\prime 3}(u), \label{fu-4}\\[4pt]
&&{R}_{3\langle 2{\bar 1}^\prime\rangle}(u)=
(u+\eta)^{-1}P^{(20) }_{{\bar 1}^\prime 2}R_{3\bar{1}^\prime }(u-\frac{1}{2}\eta) R_{32}(u+\eta)P^{(20)}_{{\bar 1}^\prime 2}\equiv R_{3{\tilde 1}^\prime}(u). \label{fu-44}
\eea
For simplicity, we denote the projected space as
$V_{\tilde 1^\prime}=V_{\langle \bar 1^\prime 2\rangle}=V_{\langle 2 \bar 1^\prime \rangle}$.
The fused $R$-matrix $R_{\tilde{1}^\prime 2}(u)$ is a $80\times 80$ one defined in the product spaces $V_{{\tilde 1}^\prime}\otimes V_2$ and
satisfies following graded Yang-Baxter equation
\bea
R_{{\tilde 1}^\prime 2}(u-v) R_{{\tilde 1}^\prime 3}(u) R_{{2}3}(v)
= R_{{2}3}(v) R_{{\tilde 1}^\prime 3}(u)R_{{\tilde 1}^\prime 2}(u-v).\label{fwusdfwa-44}
\eea

A remarkable fact is that after taking the correspondences
\bea
|\phi_i\rangle\longrightarrow|\psi_i\rangle,\quad |\tilde\phi_i\rangle\longrightarrow|\tilde\psi_i\rangle, \quad i=1,\cdots,20,\label{vec-corresp}
\eea
the two fused $R$-matrices $R_{\tilde 1 2}(u)$ given by (\ref{fu-2}) and $R_{{\tilde 1}^\prime 2}(u)$ given by (\ref{fu-4}) are identical,
\bea
R_{\tilde 1 2}(u)=R_{{\tilde 1}^\prime 2}(u),\label{peri-iden}
\eea
which allows us to close the recursive fusion processe.

The fusion of $R_{\bar{1}^\prime 2}(u)$ in the quantum space is carried out by the projector $P_{23}^{(8)}$, and the resulted fused $R$-matrix is
\bea
R_{{\bar 1}^\prime \langle 23\rangle}(u)= (u+\eta)^{-1}P_{23}^{(8)}R_{{\bar 1}^\prime 3}(u-\frac{1}{2}\eta)R_{{\bar 1}^\prime 2}(u+\frac{1}{2}\eta)P_{23}^{(8)}\equiv  R_{{\bar 1}^\prime \bar 2}(u), \eea
which is a $64\times 64$ matrix defined in the space $V_{\bar{1}^\prime }\otimes V_{\bar 2}$ and satisfies the graded Yang-Baxter equation
\bea
R_{{\bar 1}^\prime\bar 2}(u-v)R_{{\bar 1}^\prime 3}(u)R_{\bar 2 3}(v)=R_{\bar 2 3}(v)R_{{\bar 1}^\prime 3}(u)R_{{\bar 1}^\prime\bar 2}(u-v),\label{sdfusdsd-22}
\eea
which will help us to find the complete set of conserved quantities.

\subsection{Operator product identities}

Now, we are ready to extend the fusion from one site to the whole system.
From the fused $R$-matrices given by (\ref{hhgg-1}), (\ref{fu-2}), (\ref{hhgg-3}) and (\ref{fu-4}), we construct the fused monodromy matrices as
\begin{eqnarray}
&&T_{\bar 0}(u)=R_{\bar 01}(u-\theta_1)R_{\bar 02}(u-\theta_2)\cdots R_{\bar 0N}(u-\theta_N), \no \\
&&T_{\bar 0^\prime}(u)=R_{\bar 0^\prime 1}(u-\theta_1)R_{\bar 0^\prime 2}(u-\theta_2)\cdots R_{\bar 0^\prime N}(u-\theta_N), \no \\
&&T_{\tilde 0}(u)=R_{\tilde 01}(u-\theta_1)R_{\tilde 02}(u-\theta_2)\cdots R_{\tilde 0N}(u-\theta_N), \no \\
&&T_{\tilde 0^\prime}(u)=R_{\tilde 0^\prime1}(u-\theta_1)R_{\tilde 0^\prime2}(u-\theta_2)\cdots R_{\tilde 0^\prime N}(u-\theta_N),\label{T6}
\end{eqnarray}
where the subscripts $\bar 0$, $\bar 0^\prime$, $\tilde 0$ and $\tilde 0^\prime $ mean the auxiliary spaces, and
the quantum spaces in all the monodromy matrices are the same. By using the graded Yang-Baxter equations (\ref{fuse-qybe1}), (\ref{sdfu-22}),
(\ref{fuse-qybe2}), (\ref{fwusdfwa-44}) and (\ref{sdfusdsd-22}), one can prove that the monodromy matrices
satisfy the graded Yang-Baxter relations
\begin{eqnarray}
&&R_{\bar 12} (u-v) T_{\bar 1}(u) T_2(v)=  T_2(v) T_{\bar 1}(u) R_{\bar 12}(u-v), \no \\
&&R_{\bar 1^\prime 2} (u-v) T_{\bar 1^\prime }(u) T_2(v)=  T_2(v) T_{\bar 1^\prime }(u) R_{\bar 1^\prime 2} (u-v), \no \\
&&R_{\bar 1^\prime \bar 2} (u-v) T_{\bar 1^\prime }(u) T_{\bar 2}(v)=  T_{\bar 2}(v) T_{\bar 1^\prime }(u) R_{\bar 1^\prime \bar 2} (u-v), \no \\
&&R_{\tilde 12} (u-v) T_{\tilde 1}(u) T_2(v)=  T_2(v) T_{\tilde 1}(u) R_{\tilde 12}(u-v), \no \\
&&R_{\tilde 1^\prime 2} (u-v) T_{\tilde 1^\prime }(u) T_2(v)=  T_2(v) T_{\tilde 1^\prime }(u) R_{\tilde 1^\prime 2} (u-v). \label{yybb4}
\end{eqnarray}
According to the property that the $R$-matrices in above equations can degenerate into
the projectors $P^{(8)}_{12}$, $\bar P^{(8)}_{12}$, $P^{(20)}_{\bar{1}2} $, $P^{(20)}_{\bar{1}^\prime 2}$ and using
the definitions (\ref{T6}),
we obtain following fusion relations among the monodromy matrices
\bea
&&P^{ (8) }_{12}T_2 (u)T_1 (u+\eta)P^{(8) }_{12}=\prod_{l=1}^N
(u-\theta_l+\eta)T_{\bar 1}(u+\frac{1}{2}\eta), \no\\[4pt]
&&\bar P^{ (8) }_{12}T_2 (u)T_1 (u-\eta)\bar P^{(8) }_{12}
=\prod_{l=1}^N
(u-\theta_l-\eta)T_{\bar 1^\prime}(u-\frac{1}{2}\eta), \no\\[4pt]
&&P^{(20) }_{2\bar{1}} T_{\bar{1}} (u+\frac{1}{2}\eta) T_2(u-\eta)P^{(20)}_{2\bar{1}}
=\prod_{l=1}^N
(u-\theta_l-\eta){T}_{\tilde 1}(u),\no\\[4pt]
&&P^{(20) }_{2\bar{1}^\prime } T_{\bar{1}^\prime } (u-\frac{1}{2}\eta)T_2(u+\eta)P^{(20)
}_{2\bar{1}^\prime }=\prod_{l=1}^N
(u-\theta_l+\eta){T}_{\tilde 1^\prime }(u).\label{fut-6}
\eea

The fused transfer matrices are defined as the super partial traces of fused monodromy matrices in the auxiliary space
\bea
{t}^{(1)}_p(u)=str_{\bar 0} T_{\bar 0}(u), \; {t}^{(2)}_p(u)=str_{\bar 0^\prime} T_{\bar 0^\prime}(u), \;
\tilde{t}^{(1)}_p(u)=str_{\tilde 0} T_{\tilde 0}(u), \; \tilde{t}^{(2)}_p(u)=str_{\tilde 0^\prime} T_{\tilde 0^\prime }(u).\no
\eea
From Eq.(\ref{fut-6}), we know that these fused transfer matrices with certain spectral difference must satisfy some
intrinsic relations. We first consider the quantity
\bea
\hspace{-1.2truecm}&&\hspace{-1.2truecm}t_p(u)t_p(u+\eta)=str_{12}\{T_1(u)T_2(u+\eta)\}\no\\[4pt]
&&\hspace{8mm}=str_{12}\{(P_{12}^{(8)}+\bar P_{12}^{(8)})T_1(u)T_2(u+\eta)(P_{12}^{(8)}+\bar P_{12}^{(8)})\}\no\\[4pt]
&&\hspace{8mm}=str_{12}\{P_{12}^{(8)}T_1(u)T_2(u+\eta)P_{12}^{(8)}\}+str_{12}\{\bar P_{12}^{(8)}\bar P_{12}^{(8)}T_1(u)T_2(u+\eta)\bar P_{12}^{(8)}\}\no\\[4pt]
&&\hspace{8mm}=str_{12}\{P_{12}^{(8)}T_1(u)T_2(u+\eta)P_{12}^{(8)}\}+str_{12}\{\bar P_{12}^{(8)}T_2(u+\eta)T_1(u)\bar P_{12}^{(8)}\bar P_{12}^{(8)}\}\no\\[4pt]
&&\hspace{8mm}=\prod_{j=1}^{N}(u-\theta_j+\eta) t_p^{(1)}(u+\frac{1}{2}\eta)+\prod_{j=1}^{N}(u-\theta_j) t_p^{(2)}(u+\frac{1}{2}\eta).\label{fui-3tan}
\eea
Here we give some remarks. Both $V_1$ and $V_2$ are the 4-dimensional auxiliary spaces.
From Eq.(\ref{fui-3tan}), we see that the 16-dimensional auxiliary space $V_{1}\otimes V_2$
can be projected into two 8-dimensional subspaces, $V_{1}\otimes V_2=V_{\langle12\rangle}\oplus V_{\langle12\rangle^\prime}$.
One is achieved by the 8-dimensional projector $P_{12}^{(8)}$ defined in the subspace $V_{\langle12\rangle}\equiv V_{\bar 1}$,
and the other is achieved by the 8-dimensional projector $\bar P_{12}^{(8)}$ defined in the subspace $V_{\langle 12\rangle^\prime}\equiv V_{\bar 1^\prime}$.
The vectors in $P_{12}^{(8)}$ and those in $\bar P_{12}^{(8)}$ constitute the complete basis of $V_{1}\otimes V_2$, and all the vectors are orthogonal,
\bea
P_{12}^{(8)}+\bar P_{12}^{(8)}=1,~~P_{12}^{(8)}\bar P_{12}^{(8)}=0.\no
\eea
From Eq.(\ref{fui-3tan}), we also know that the product of two transfer matrices with fixed spectral difference can be written as the summation of
two fused transfer matrices $t_p^{(1)}(u)$ and $ t_p^{(2)}(u)$.
At the point of $u=\theta_j-\eta$, the coefficient of the fused transfer matrix $ t_p^{(1)}(u)$  is zero, while at
the point of $u=\theta_j$, the coefficient of the fused transfer matrix $ t_p^{(2)}(u)$  is zero.
Therefore, at these points, only one of them has the contribution.

Motivated by Eq.(\ref{fut-6}), we also consider the quantities
\bea
\hspace{-0.8truecm}&&\hspace{-0.8truecm} t_p^{(1)}(u+\frac{1}{2}\eta)t_p(u-\eta)=str_{\bar 12}\{(P_{2\bar 1}^{(20)}+\tilde P_{2\bar 1}^{(12)})T_{\bar 1}(u+\frac{1}{2}\eta)T_2(u-\eta)(P_{2\bar 1}^{(20)}+\tilde P_{2\bar 1}^{(12)})\}\no\\[4pt]
&&=str_{\bar 12}\{P_{2\bar 1}^{(20)}T_{\bar 1}(u+\frac{1}{2}\eta)T_2(u-\eta)P_{2\bar 1}^{(20)}\}+str_{\bar 12}\{\tilde P_{2\bar 1}^{(12)}T_{\bar 1}(u+\frac{1}{2}\eta)T_2(u-\eta)\tilde P_{2\bar 1}^{(12)}\}\no\\[4pt]
&&=\prod_{j=1}^{N}(u-\theta_j-\eta)\tilde t_p^{(1)}(u)+\prod_{j=1}^{N}(u-\theta_j)\bar{t}_p^{(1)}(u), \label{fui-3tan-1}\\
\hspace{-0.8truecm}&&\hspace{-0.8truecm} t_p^{(2)}(u-\frac{1}{2}\eta)t_p(u+\eta)=str_{\bar 1^\prime 2}\{(P_{2\bar 1^\prime }^{(20)}
+\tilde P_{2\bar 1^\prime }^{(12)})T_{\bar 1^\prime }(u-\frac{1}{2}\eta)T_2(u+\eta)(P_{2\bar 1^\prime }^{(20)}+\tilde P_{2\bar 1^\prime }^{(12)})\}\no\\[4pt]
&&=str_{\bar 1^\prime 2}\{P_{2\bar 1^\prime }^{(20)}T_{\bar 1^\prime }(u-\frac{1}{2}\eta)T_2(u+\eta)P_{2\bar 1^\prime }^{(20)}\}
+str_{\bar 1^\prime 2}\{\tilde P_{2\bar 1^\prime }^{(12)}T_{\bar 1^\prime }(u-\frac{1}{2}\eta)T_2(u+\eta)\tilde P_{2\bar 1^\prime }^{(12)}\}\no\\[4pt]
&&=\prod_{j=1}^{N}(u-\theta_j+\eta)\tilde t_p^{(2)}(u)+\prod_{j=1}^{N}(u-\theta_j)\bar{t}_p^{(2)}(u).\label{fui-3tan-2}
\eea
During the derivation, we have used the relations
\bea
P_{2\bar 1}^{(20)}+\tilde P_{2\bar 1}^{(12)}=1,~~P_{2\bar 1}^{(20)}\tilde P_{2\bar 1}^{(12)}=0, ~~
P_{2\bar 1^\prime}^{(20)}+\tilde P_{2\bar 1^\prime}^{(12)}=1,~~P_{2\bar 1^\prime}^{(20)}\tilde P_{2\bar 1^\prime}^{(12)}=0.\no
\eea
From Eq.(\ref{fui-3tan-1}), we see that the 32-dimensional auxiliary space $V_{\bar 1}\otimes V_2$
can be projected into a 20-dimensional subspace $V_{\langle \bar 12\rangle}\equiv V_{\tilde 1}$ by the projector $P_{\bar 12}^{(20)}$
and a 12-dimensional subspace $V_{\overline{\langle \bar 12\rangle}}$ by the projector $\tilde P_{\bar 12}^{(12)}$, $V_{\bar 1}\otimes V_2=V_{\langle \bar 12\rangle} \oplus
V_{\overline{\langle \bar 12\rangle}}$.
The vectors in $P_{\bar 12}^{(20)}$ and $\tilde P_{\bar 12}^{(12)}$ are the complete and orthogonal
basis.
Eq.(\ref{fui-3tan-1}) also gives that the quantity $t_p^{(1)}(u+\frac{1}{2}\eta)t_p(u-\eta)$
is the summation of two new fused transfer matrices $\tilde t_p^{(1)}(u)$ and $\bar{t}_p^{(1)}(u)$ with some coefficients.
In Eq.(\ref{fui-3tan-2}), the 32-dimensional auxiliary space $V_{\bar 1^\prime}\otimes V_2$ is
projected into a 20-dimensional and a 12-dimensional subspaces by the operators $P_{\bar 1^\prime 2}^{(20)}$ and $\tilde P_{\bar 1^\prime 2}^{(12)}$, respectively.
Thus the quantity $t_p^{(2)}(u-\frac{1}{2}\eta)t_p(u+\eta)$ is the summation of two fused transfer matrices $\tilde t_p^{(2)}(u)$ and $\bar{t}_p^{(2)}(u)$ with some coefficients.
At the point of $u=\theta_j-\eta$, the coefficient of $\tilde t_p^{(1)}(u)$ in Eq.(\ref{fui-3tan-1})
and that of $\tilde t_p^{(2)}(u)$ in Eq.(\ref{fui-3tan-1}) are zero. While at the point of $u=\theta_j$, the coefficient of $\bar{t}_p^{(1)}(u)$ in Eq.(\ref{fui-3tan-1})
and that of $\bar{t}_p^{(2)}(u)$ in Eq.(\ref{fui-3tan-2}) are zero.
Here, the explicit forms of $\tilde P_{\bar 12}^{(12)}$, $\tilde P_{\bar 1^\prime 2}^{(12)}$, $\bar{t}_p^{(1)}(u)$ and $\bar{t}_p^{(2)}(u)$ are omitted
because we donot use them.

Combining the above analysis, we obtain the operator product identities of the transfer matrices at the fixed points as
\bea && t_p(\theta_j)t_p (\theta_j+\eta)=\prod_{l=1}^N
(\theta_j-\theta_l+\eta) t^{(1)}_p(\theta_j+\frac{1}{2}\eta),\label{futp-4-1} \\[4pt]
&& t_p(\theta_j)t_p (\theta_j-\eta)=\prod_{l=1}^N
(\theta_j-\theta_l-\eta) t^{(2)}_p(\theta_j-\frac{1}{2}\eta),\label{futp-4-2} \\[4pt]
&& t^{(1)}_p(\theta_j+\frac{1}{2}\eta)t_p (\theta_j-\eta)=\prod_{l=1}^N
(\theta_j-\theta_l-\eta)\tilde t_{p}^{(1)}(\theta_j),\label{futp-4-3}\\[4pt]
&& t^{(2)}_p(\theta_j-\frac{1}{2}\eta)t_p (\theta_j+\eta)=\prod_{l=1}^N
(\theta_j-\theta_l+\eta)\tilde t_{p}^{(2)}(\theta_j), \quad j=1, \cdots, N.\label{futp-4-4}
\eea
From the property (\ref{peri-iden}), we obtain that the fused transfer matrices $\tilde{t}^{(1)}_p(u)$ and $\tilde{t}^{(2)}_p(u)$
are equal
\bea
\tilde{t}^{(1)}_p(u)=\tilde{t}^{(2)}_p(u). \label{futp-6}
\eea
With the help of  Eqs. (\ref{futp-6}), (\ref{futp-4-3}) and (\ref{futp-4-4}),
we can obtain the constraint among $t_p(u)$, $ t^{(1)}_p(u)$ and $ t^{(2)}_p(u)$,
\bea
 t^{(1)}_p (\theta_j+\frac{1}{2}\eta) t_p(\theta_j-\eta)
=\prod_{l=1}^N\frac{\theta_j-\theta_l-\eta}{\theta_j-\theta_l+\eta} t^{(2)}_p (\theta_j-\frac{1}{2}\eta) t_p(\theta_j+\eta).\label{peri-ope-3}
\eea
Then Eqs.(\ref{futp-4-1}), (\ref{futp-4-2}) and (\ref{peri-ope-3}) constitute the closed recursive fusion relations.
From the definitions, we know that the transfer matrices $t_p(u)$, ${t}^{(1)}_p(u)$ and ${t}^{(2)}_p(u)$
are the operator polynomials of $u$ with degree $N-1$. Then, the $3N$ conditions (\ref{futp-4-1}), (\ref{futp-4-2}) and (\ref{peri-ope-3})
are sufficient to solve them.

From the graded Yang-Baxter relations (\ref{yybb4}), the transfer matrices $t_p(u)$, ${t}^{(1)}_p(u)$ and ${t}^{(2)}_p(u)$
commutate with each other, namely,
\bea
[t_p(u),{t}^{(1)}_p(u)]=[t_p(u),{t}^{(2)}_p(u)]=[{t}^{(1)}_p(u),{t}^{(2)}_p(u)]=0.
\eea
Therefore, they have common eigenstates and can be diagonalized simultaneously.
Let $|\Phi\rangle$ be a common eigenstate. Acting the transfer matrices on this eigenstate, we have
\bea
t_p(u)|\Phi\rangle=\Lambda_p(u)|\Phi\rangle,\quad
t_p^{(1)}(u)|\Phi\rangle= \Lambda_p^{(1)}(u)|\Phi\rangle,\quad
t_p^{(2)}(u)|\Phi\rangle=\Lambda_p^{(2)}(u)|\Phi\rangle,\no
\eea
where $\Lambda_p(u)$, ${\Lambda}^{(1)}_p(u)$ and ${\Lambda}^{(2)}_p(u)$ are the eigenvalues of
$t_p(u)$, ${t}^{(1)}_p(u)$ and ${t}^{(2)}_p(u)$, respectively. Meanwhile, acting the operator product identities (\ref{futp-4-1}),
(\ref{futp-4-2}) and (\ref{peri-ope-3}) on the state $|\Phi\rangle$, we have the functional relations among these eigenvalues
\bea && \Lambda_p(\theta_j)\Lambda_p (\theta_j+\eta)=\prod_{l=1}^N
(\theta_j-\theta_l+\eta){\Lambda}^{(1)}_p(\theta_j+\frac{1}{2}\eta),\no \\[4pt]
&& \Lambda_p(\theta_j)\Lambda_p (\theta_j-\eta)=\prod_{l=1}^N
(\theta_j-\theta_l-\eta){\Lambda}^{(2)}_p(\theta_j-\frac{1}{2}\eta),\no \\[4pt]
&& \Lambda^{(1)}_p (\theta_j+\frac{1}{2}\eta) \Lambda_p(\theta_j-\eta)=\prod_{l=1}^N\frac{\theta_j-\theta_l-\eta}{\theta_j-\theta_l+\eta}
\Lambda^{(2)}_p (\theta_j-\frac{1}{2}\eta) \Lambda_p(\theta_j+\eta),\label{futpl-3}
\eea
where $j=1,2,\cdots N$. Because the eigenvalues $\Lambda_p(u)$, ${\Lambda}^{(1)}_p(u)$ and ${\Lambda}^{(2)}_p(u)$
are the polynomials of $u$ with degree $N-1$, the above $3N$ conditions (\ref{futpl-3}) can determine these eigenvalues completely.

\subsection{$T-Q$ relations}

Let us introduce the $z$-functions
\begin{eqnarray}
z_p^{(l)}(u)=\left\{
\begin{array}{ll}
\displaystyle(-1)^{p(l)}Q^{(0)}_p(u)\frac{Q_p^{(l-1)}(u+\eta)Q_p^{(l)}(u-\eta)}{Q_p^{(l)}(u)Q_p^{(l-1)}(u)}, &l=1,2,\\[6mm]
\displaystyle(-1)^{p(l)}Q^{(0)}_p(u)\frac{Q_p^{(l-1)}(u-\eta)Q_p^{(l)}(u+\eta)}{Q_p^{(l)}(u)Q_p^{(l-1)}(u)}, &l=3,4,\end{array}
\right.
\end{eqnarray}
where the $Q$-functions are
\bea
&&Q_p^{(0)}(u)=\prod_{l=1}^{N}(u-\theta_j),\quad
Q^{(m)}_p(u)=\prod_{j=1}^{L_m}(u-\lambda_j^{(m)}), \quad m=1, 2, 3,\quad Q_p^{(4)}(u)=1,\no
\eea
and $\{L_m|m=1,2,3\}$ are the numbers of the Bethe roots $\{\lambda_j^{(m)}\}$.

According to the closed functional relations (\ref{futpl-3}), we construct the eigenvalues of the transfer matrices in terms of the homogeneous $T-Q$ relations
 \bea &&\Lambda_p (u)=\sum_{l=1}^{4}z_p^{(l)}(u),
\no\\[4pt]
&&\Lambda_p^{(1)}(u)=\Big[Q_p^{(0)}(u+\frac{1}{2}\eta)\Big]^{-1}\Big\{\sum_{l=1}^{2}z_p^{(l)}(u+\frac{1}{2}\eta)z_p^{(l)}(u-\frac{1}{2}\eta)\no\\
&&~~~~~~~~~~~~~~~+\sum_{l=2}^{4}
\sum_{m=1}^{l-1}z_p^{(l)}(u+\frac{1}{2}\eta)z_p^{(m)}(u-\frac{1}{2}\eta)\Big\},
\no\\[4pt]
&&\Lambda_p^{(2)}(u)=\Big[Q_p^{(0)}(u-\frac{1}{2}\eta)\Big]^{-1}\Big\{\sum_{l=3}^{4}z_p^{(l)}(u+\frac{1}{2}\eta)z_p^{(l)}(u-\frac{1}{2}\eta)\no\\[4pt]
&&~~~~~~~~~~~~~~~+\sum_{l=2}^{4}
\sum_{m=1}^{l-1}z_p^{(l)}(u-\frac{1}{2}\eta)z_p^{(m)}(u+\frac{1}{2}\eta)\Big\}.\label{ep-3}
\eea
The regularities of the eigenvalues $\Lambda_p(u)$, $\Lambda_p^{(1)}(u)$ and $\Lambda_p^{(2)}(u)$ give rise to the constraints that the Bethe roots
$\{\lambda_j^{(m)}\}$ should satisfy the Bethe ansatz equations (BAEs)
\bea &&\frac{Q_p^{(0)}(\lambda_j^{(1)}+\eta)}{Q_p^{(0)}(\lambda_j^{(1)})}=-\frac{Q_p^{(1)}(\lambda_j^{(1)}+\eta)Q_p^{(2)}(\lambda_j^{(1)}-\eta)}
{Q_p^{(2)}(\lambda_j^{(1)})Q_p^{(1)}(\lambda_j^{(1)}-\eta)},~~j=1,\cdots,L_1,
\no\\
&&\frac{Q_p^{(1)}(\lambda_j^{(2)}+\eta)}{Q_p^{(1)}(\lambda_j^{(2)})}=\frac{Q_p^{(3)}(\lambda_j^{(2)})}{Q_p^{(3)}(\lambda_j^{(2)})},~~j=1,\cdots,L_2,
\no\\
&&\frac{Q_p^{(2)}(\lambda_j^{(3)}-\eta)}{Q_p^{(2)}(\lambda_j^{(3)})}=-\frac{Q_p^{(3)}(\lambda_j^{(3)}-\eta)}{Q_p^{(3)}(\lambda_j^{(3)}+\eta)},~~j=1,\cdots,L_3. \label{BAE-period-3}\eea

We have verified that the above BAEs indeed guarantee all the $T-Q$ relations (\ref{ep-3}) are polynomials and
satisfy the functional relations (\ref{futpl-3}). Therefore, we arrive at the conclusion
that $\Lambda_p(u)$, $\Lambda_p^{(1)}(u)$ and $\Lambda_p^{(2)}(u)$ given by (\ref{ep-3}) are indeed the eigenvalues of
the transfer matrices $t_p(u)$, ${t}^{(1)}_p(u)$, ${t}^{(2)}_p(u)$, respectively.
The eigenvalues of the Hamiltonian (\ref{peri-Ham}) are
\begin{eqnarray}
E_p= \frac{\partial \ln \Lambda_p(u)}{\partial
u}|_{u=0,\{\theta_j\}=0}.
\end{eqnarray}

\section{$SU(2|2)$ model with off-diagonal boundary reflections}
\setcounter{equation}{0}

\subsection{Boundary integrability}

In this section, we consider the system with open boundary conditions.
The boundary reflections are characterized by the reflection matrix $K^-(u)$ at one side and $K^+(u)$ at the other side.
The integrability requires that $K^-(u)$ satisfies the graded reflection equation (RE) \cite{Che84, Bra98}
\begin{equation}
 R _{12}(u-v){K^{-}_{  1}}(u)R _{21}(u+v) {K^{-}_{2}}(v)=
 {K^{-}_{2}}(v)R _{12}(u+v){K^{-}_{1}}(u)R _{21}(u-v),
 \label{r1}
 \end{equation}
while $K^+(u)$ satisfies the graded dual RE
\begin{eqnarray}
R_{12}(v-u)K_1^+(u)R_{21}(-u-v)K_2^+(v)=K_2^+(v)R_{12}(-u-v)K_1^+(u)R_{21}(v-u).
 \label{r2}
 \end{eqnarray}
The general solution of reflection matrix $K_0^{-}(u)$ defined in the space $V_0$ satisfying the graded RE (\ref{r1}) is
\bea
K_0^{-}(u)=\xi+uM,\quad M=\left(\begin{array}{cccc}1 &c_1&0&0\\[6pt]
c_2&-1 &0&0\\[6pt]
0&0 &-1 &c_3\\[6pt]
0&0&c_4&1 \end{array}\right), \label{K-matrix-1}\eea
and the dual reflection matrix $K^+(u)$ can be obtained by the mapping
\begin{equation}
K_0^{ +}(u)=K_0^{ -}(-u)|_{\xi,c_i\rightarrow
\tilde{\xi},\tilde{c}_i }, \label{K-matrix-2}
\end{equation}
where the $\xi$, $\tilde{\xi}$ and $\{c_i, \tilde{c}_i |i=1,\cdots,4\}$
are the boundary parameters which describe
the boundary interactions, and the integrability requires
\bea
c_1c_2=c_3c_4,\quad
\tilde{c}_1\tilde{c}_2=\tilde{c}_3\tilde{c}_4.\no
\eea

The reflection matrices (\ref{K-matrix-1}) and (\ref{K-matrix-2}) have the off-diagonal elements,
thus the numbers of ``quasi-particles" with different intrinsic degrees of freedom are not conserved during the reflection processes.
Meanwhile, the $K^-(u)$ and $K^+(u)$ are not commutative,
$[K^-(u),K^+(v)]$ $\neq 0$, which means that they cannot be diagonalized simultaneously.
Thus it is quite hard to derive the exact solutions of the system via the conventional Bethe ansatz because of the
absence of a proper reference state. We will develop the graded nested ODBA to solve the system exactly.

For the open case, besides the standard ``row-to-row" monodromy matrix $T_0(u)$ specified by (\ref{T1}), one needs to
consider the reflecting monodromy matrix
\begin{eqnarray}
\hat{T}_0 (u)=R_{N0}(u+\theta_N)\cdots R_{20}(u+\theta_{2}) R_{10}(u+\theta_1),\label{Tt11}
\end{eqnarray}
which satisfies the graded Yang-Baxter relation
\begin{eqnarray}
R_{ 12} (u-v) \hat T_{1}(u) \hat T_2(v)=\hat  T_2(v) \hat T_{ 1}(u) R_{12} (u-v)\label{haishi0}.
\end{eqnarray}
The transfer matrix $t(u)$ is defined as
\begin{equation}
t(u)= str_0 \{K_0^{ +}(u)T_0 (u) K^{ -}_0(u)\hat{T}_0 (u)\}\label{tru}.
\end{equation}
The graded Yang-Baxter relations (\ref{ybe}), (\ref{haishi0}) and reflection equations (\ref{r1}), (\ref{r2})
lead to the fact that the transfer matrices with different spectral parameters commutate with each other, $[t(u), t(v)]=0$. Therefore, $t(u)$ serves
as the generating function of all the conserved quantities and the system is integrable.
The model Hamiltonian with open boundary condition can be written out in terms of transfer matrix (\ref{tru}) as
\begin{eqnarray}
H=\frac{1}{2}\frac{\partial \ln t(u)}{\partial
u}|_{u=0,\{\theta_j\}=0}. \label{hh}
\end{eqnarray}
The hermiticity of Hamiltonian (\ref{hh}) further requires $c_1=c_2^{*}$ and $c_3=c_4^{*}$.

\subsection{Fused reflection matrices}

In order to solve the eigenvalue problem of the  transfer matrix (\ref{tru}), we should study the fusion of boundary reflection matrices \cite{Mez92, Zho96}. The main idea of the fusion for reflection matrices associated with a supersymmetric model is expressed in Appendix A. Focusing on the supersymmetric $SU(2|2)$ model with the boundary reflection matrices (\ref{K-matrix-1}) and (\ref{K-matrix-2}), we can take fusion according to Eqs.(\ref{oled-3})-(\ref{oled-4}) or
(\ref{oled-13})-(\ref{oled-14}).
The two 8-dimensional fusion associated with  the super projectors $P_{12}^{(8)}$ (\ref{1-project}) and $\bar P_{12}^{(8)}$ (\ref{2-project}) gives
\bea
&&
{K}^{-}_{\bar 1}(u)=(u+\frac{1}{2}\eta)^{-1}P_{21}^{(8)}K_1^{-}(u-\frac{1}{2}\eta)R_{21}(2u)K_2^{-}(u+\frac{1}{2}\eta)P_{12}^{(8)},\no\\[4pt]
&&
{K}^{+}_{\bar 1}(u)=(u-\frac{1}{2}\eta)^{-1}P_{12}^{(8)}K_2^+(u+\frac{1}{2}\eta)R_{12}(-2u)K_1^+(u-\frac{1}{2}\eta)P_{21}^{(8)},\no\\[4pt]
&& {K}^{-}_{\bar 1^\prime }(u)=(u-\frac{1}{2}\eta)^{-1}\bar P_{21}^{(8)}K_1^{-}(u+\frac{1}{2}\eta)R_{21}(2u)K_2^{-}
(u-\frac{1}{2}\eta)\bar P_{12}^{(8)},\no\\[4pt]
&& K^{+}_{\bar 1^\prime}(u)=(u+\frac{1}{2}\eta)^{-1}\bar P_{12}^{(8)}K_2^{+}(u-\frac{1}{2}\eta)
R_{12}(-2u)K_1^{+}(u+\frac{1}{2}\eta)\bar P_{21}^{(8)}.\label{open-k4}
 \eea
By specific calculation, we know that all the fused $K$-matrices are the $8\times8$ ones and their matric elements are the polynomials of $u$ with maximum degree two.
The fused reflection $K$-matrices (\ref{open-k4}) satisfy the resulting graded reflection equations. We can further use
the reflection matrices $K_{\bar 1}^{\pm}(u)$ [or $K_{\bar 1^\prime }^{\pm}(u)$] and $K_2^{\pm}(u)$
to obtain the $20$-dimensional projector $ P_{{\bar 1}2}^{(20)}$ (\ref{cjjc}) [or $P_{{\bar 1}^\prime 2}^{(20)}$ (\ref{xiao1})]. The
resulted new fused reflection matrices are
\bea && {K}^{-}_{\tilde 1}(u)=(u-
\eta)^{-1}
P_{2{\bar1}}^{(20)} K_{\bar{1}}^{-}(u+\frac{1}{2}\eta)R_{2\bar 1}(2u-\frac{1}{{2}}\eta)K_{2}^{-}(u-\eta)P_{{\bar 1}2}^{(20)}, \no \\[4pt]
&& {K}^{+}_{\tilde 1}(u)=(2u+\eta)^{-1}
P_{{\bar 1}2}^{(20)} K_{2}^{+}(u-\eta)R_{{\bar 1}2}(-2u+\frac{1}{{2}}\eta) K_{\bar{1}}^{+}(u+\frac{1}{2}\eta)P_{2{\bar 1}}^{(20)}, \no \\[4pt]
&& {K}^{-}_{\tilde 1^\prime }(u)=(u+
\eta)^{-1}
P_{2{\bar1^\prime }}^{(20)} K_{\bar{1}^\prime }^{-}(u-\frac{1}{2}\eta)R_{2\bar 1^\prime}(2u+\frac{1}{{2}}\eta) K_{2}^{-}(u+\eta)P_{{\bar 1}^\prime 2}^{(20)},\no \\[4pt]
&& {K}^{+}_{\tilde 1^\prime }(u)=(2u-\eta)^{-1}
P_{{\bar 1}^\prime 2}^{(20)} K_{2}^{+}(u+\eta)R_{{\bar 1}^\prime 2}(-2u-\frac{1}{{2}}\eta) K_{\bar{1}^\prime }^{+}(u-\frac{1}{2}\eta)P_{2{\bar 1^\prime }}^{(20)}.\label{fuseref4} \eea
It is easy to check that the fused reflection matrices (\ref{fuseref4}) are the $20\times 20$ ones where the
matric elements are polynomials of $u$ with maximum degree three.
Moreover, keeping the correspondences (\ref{vec-corresp}) in mind, we have the important relations that the fused
reflection matrices defined in the projected subspace $V_{ \tilde 1}$ and that defined in the projected subspace $V_{ \tilde 1^\prime}$ are equal
\bea
{K}^{-}_{\tilde 1}(u)={K}^{-}_{\tilde 1^\prime }(u), \quad {K}^{+}_{\tilde 1}(u)={K}^{+}_{\tilde 1^\prime }(u), \label{k-iden}
\eea
which will be used to close the fusion processes with boundary reflections.
\subsection{Operator production identities}

For the model with open boundary condition, besides the fused monodromy matrices (\ref{T6}), we also need the fused reflecting monodromy matrices, which are
constructed as
\begin{eqnarray}
&&\hat{T}_{\bar 0}(u)=R_{ N\bar 0}(u+\theta_N)\cdots R_{2\bar 0}(u+\theta_2)R_{1\bar 0}(u+\theta_1), \no \\[4pt]
&&\hat{T}_{\bar 0^\prime}(u)=R_{N\bar 0^\prime}(u+\theta_N)\cdots R_{2\bar 0^\prime}(u+\theta_2)R_{1\bar 0^\prime}(u+\theta_1).\label{openT6}
\end{eqnarray}
The fused reflecting monodromy matrices satisfy the graded Yang-Baxter relations
\begin{eqnarray}
&&R_{1\bar 2} (u-v) \hat{T}_1(u) \hat{ T}_{\bar 2}(v) = \hat{ T}_{\bar 2}(v) \hat{T}_1(u) R_{1\bar 2} (u-v), \no \\[4pt]
&&R_{1\bar 2^\prime} (u-v) \hat{T}_1(u) \hat{T}_{\bar 2^\prime}(v) = \hat{ T}_{\bar 2^\prime}(v) \hat{T}_1(u) R_{1\bar 2^\prime} (u-v), \no \\[4pt]
&&R_{\bar 1\bar 2^\prime} (u-v) \hat{T}_{\bar 1}(u) \hat{T}_{\bar 2^\prime}(v) = \hat{ T}_{\bar 2^\prime}(v) \hat{T}_{\bar 1}(u) R_{\bar 1\bar 2^\prime} (u-v).\label{yyBB222}
\end{eqnarray}
The fused transfer matrices are defined as
\bea
&&t^{(1)}(u)= str_{\bar 0}\{K^{+}_{\bar{0}}(u)  T_{\bar 0}(u)  K^{-}_{\bar{0}}(u) \hat{T}_{\bar 0}(u)\},\no \\[4pt]
&&t^{(2)}(u)= str_{\bar 0^\prime}\{K^{+}_{\bar{0}^\prime}(u)   T_{\bar 0^\prime}(u)    K^{-}_{\bar{0}^\prime}(u) \hat{ T}_{\bar 0^\prime}(u)\}.\label{openTransfer-5}\eea

Using the method we have used in the periodic case, we can obtain the operator product identities among the fused transfer matrices as

\bea && t (\pm\theta_j)t (\pm\theta_j+\eta)=-\frac{1}{
4} \frac{(\pm\theta_j)(\pm\theta_j+\eta)
}{(\pm\theta_j+\frac{1}{{2}}\eta)^2}\nonumber\\[4pt]
&&\hspace{20mm}\times\prod_{l=1}^N
(\pm\theta_j-\theta_l+\eta)(\pm\theta_j+\theta_l+\eta) t^{(1)}(\pm\theta_j+\frac{1}{2}\eta),\label{openident1} \\[4pt]
&& t (\pm\theta_j)t (\pm\theta_j-\eta)=-\frac{1}{
4} \frac{(\pm\theta_j)(\pm\theta_j-\eta)
}{(\pm\theta_j-\frac{1}{{2}}\eta)^2}\nonumber\\[4pt]
&&\hspace{20mm}\times\prod_{l=1}^N
(\pm\theta_j-\theta_l-\eta)(\pm\theta_j+\theta_l-\eta) t^{(2)}(\pm\theta_j-\frac{1}{2}\eta),\label{openident2}\\[4pt]
&&
t (\pm\theta_j-\eta){ t}^{(1)}(\pm\theta_j+\frac{1}{{2}}\eta)=\frac{(\pm\theta_j+\frac{1}{2}\eta)^2(\pm\theta_j-\eta)}{(\pm\theta_j+\eta)
(\pm\theta_j-\frac{1}{{2}}\eta)^2}\no\\[4pt]&&~~~~~\times
 \prod_{l=1}^N \frac{(\pm\theta_j-\theta_l-\eta)(\pm\theta_j+\theta_l-\eta) }{(\pm\theta_j-\theta_l+\eta)(\pm\theta_j+\theta_l+\eta)} t (\pm\theta_j+\eta){t}^{(2)}(\pm\theta_j-\frac{1}{{2}}\eta).\label{openident3}
\eea
The proof of the above operator identities is given in Appendix B.

From the definitions, we know that the transfer matrix $t(u)$ is a operator polynomial of $u$ with degree $2N+2$ while
the fused ones ${t}^{(1)}(u)$ and ${t}^{(2)}(u)$ are the operator polynomials of $u$ both with degree $2N+4$.
Thus they can be completely determined by $6N+13$ independent conditions.
The recursive fusion relations (\ref{openident1}), (\ref{openident2}) and (\ref{openident3}) gives $6N$ constraints and we still need 13 ones, which can be
achieved by analyzing the values of transfer matrices at some special points. After some direct calculation, we have
\bea
&& t(0)=0,\quad {t}^{(1)}(0)=0,\quad {t}^{(2)}(0)=0, \quad {t}^{(1)}(\frac{\eta}{2})=-2\xi \tilde{\xi} t(\eta), \no \\[4pt]
&& {t}^{(1)}(-\frac{\eta}{2})=-2\xi \tilde{\xi} t(-\eta), \quad
{t}^{(2)}(\frac{\eta}{2})=2\xi \tilde{\xi} t(\eta), \quad {t}^{(2)}(-\frac{\eta}{2})=2\xi \tilde{\xi} t(-\eta),\no \\[4pt]
&& \frac{\partial {t}^{(1)}(u)}{\partial u}|_{u=0}+ \frac{\partial {t}^{(2)}(u)}{\partial u}|_{u=0}=0. \label{specialvalue4}
\eea
Meanwhile, the asymptotic behaviors of $t(u)$, $ t^{(1)}(u)$ and $ t^{(2)}(u)$ read
\bea
&& t(u)|_{u\rightarrow\infty}=-[c_1\tilde{c}_2+\tilde{c}_1c_2-c_3\tilde{c}_4-\tilde{c}_3c_4] u^{2N+2}\times {\rm id}
-\eta \hat U u^{2N+1}+\cdots, \no \\[4pt]
&& {t}^{(1)}(u)|_{u\rightarrow\infty}=-4\{2[c_3c_4\tilde{c}_3\tilde{c}_4-\tilde{c_3}c_4-c_3\tilde{c}_4-1]+(1+c_1\tilde{c}_2)^2+(1+\tilde{c_1}c_2)^2\no\\[4pt]
&&\hspace{30mm}-(c_1\tilde{c}_2+\tilde{c}_1c_2)(c_3\tilde{c}_4+\tilde{c}_3c_4)\}u^{2N+4}\times{\rm id}
 -4\eta\hat Q u^{2N+3}+\cdots, \no \\[4pt]
&& {t}^{(2)}(u)|_{u\rightarrow\infty}=-4\{2[c_1c_2\tilde{c}_1\tilde{c}_2-\tilde{c}_1c_2-c_1\tilde{c}_2-1]+(1+c_3\tilde{c}_4)^2+(1+\tilde{c}_3c_4)^2\no\\[4pt]
&&\hspace{30mm}-(c_1\tilde{c}_2+\tilde{c}_1c_2)(c_3\tilde{c}_4)+\tilde{c}_3c_4\}u^{2N+4}\times{\rm id}
+\cdots.\label{openasym3}
\eea
Here we find that the operator $\hat{U}$ related to the coefficient of transfer matrix $t(u)$ with degree $2N+1$ is given by
\bea
 \hat U= \sum_{i=1}^{N}\hat U_i=\sum_{i=1}^{N}(M_i \tilde{M}_i+\tilde{M}_i M_i),\label{openasym5}
\eea
where $M_i$ is given by (\ref{K-matrix-1}), $\tilde M_i$ is determined by (\ref{K-matrix-2}) and the operator $\hat U_i$ is
\bea
\hat U_i=\left(
           \begin{array}{cccc}
                 2+c_1\tilde{c}_2+\tilde{c}_1c_2 & 0 & 0 & 0 \\
                 0 & 2+c_1\tilde{c}_2+\tilde{c}_1c_2 & 0 & 0 \\
                 0 & 0 & 2+c_3\tilde{c}_4+\tilde{c}_3c_4 & 0 \\
                 0 & 0 & 0 & 2+c_3\tilde{c}_4+\tilde{c}_3c_4 \\
                   \end{array}
     \right)_i.
\eea
We note that $\hat U_i$ is the operator defined in the $i$-th physical space $V_i$ and can be expressed by a diagonal matrix with constant elements.
The summation of $\hat U_i$ in Eq.(\ref{openasym5}) is the direct summation and the representation matrix of operator $\hat U$ is also a diagonal one with constant elements. Moreover, we find that the operator $\hat{Q}$ related to the coefficient of the fused transfer matrix ${t}^{(1)}(u)$ with degree $ 2N+3$ is given by
\bea
\hat Q=\sum_{i=1}^{N}\hat Q_i,\label{openasym4}
\eea
where the operator $\hat Q_i$ is defined in $i$-th physical space $V_i$ with the matrix form of
\bea
&&\hat Q_i=\left(
  \begin{array}{cccc}
    \alpha & 0 & 0 & 0 \\
    0 & \alpha & 0 & 0 \\
    0 & 0 & \beta & 0 \\
    0 & 0 & 0 & \beta \\
  \end{array}
\right)_i, \no \\[4pt]
&& \alpha=2-2\tilde{c}_1\tilde{c}_2+4c_1\tilde{c}_2+(c_1\tilde{c}_2)^2+4\tilde{c}_1c_2-2c_1c_2+(\tilde{c}_1c_2)^2,\no \\[4pt]
&& \beta=2-2\tilde{c}_3\tilde{c}_4-(c_1\tilde{c}_2)^2-(\tilde{c_1}c_2)^2-4c_1c_2\tilde{c}_1\tilde{c}_2+4c_3\tilde{c}_4+2c_1\tilde{c}_2c_3\tilde{c}_4\no\\[4pt]
&&\hspace{10mm}+2\tilde{c}_1c_2c_3\tilde{c}_4
+4\tilde{c}_3c_4+2c_1\tilde{c}_2\tilde{c}_3c_4+2\tilde{c}_1c_2\tilde{c}_3c_4-2c_3c_4.\no
\eea
Again, the operator $\hat Q_i$ is a diagonal matrix with constant elements and the summation of $\hat Q_i$ in Eq.(\ref{openasym4}) is the direct summation.

So far, we have found out the $6N+13$ relations (\ref{openident1}), (\ref{openident2}), (\ref{openident3}),
(\ref{specialvalue4})-(\ref{openasym4}), which allow us to determine the eigenvalues of the transfer matrices $t(u)$, ${t}^{(1)}(u)$ and $ t^{(2)}(u)$.

\subsection{Functional relations}

From the graded Yang-Baxter relations (\ref{yybb4}), (\ref{yyBB222}) and graded reflection equations (\ref{r1}) (\ref{r2}),
one can prove that the transfer matrices $t(u)$, ${t}^{(1)}(u)$ and ${t}^{(2)}(u)$
commutate with each other, namely,
\bea
[t(u), {t}^{(1)}(u)]=[t(u), {t}^{(2)}(u)]=[ {t}^{(1)}(u), {t}^{(2)}(u)]=0.\label{opencom}
\eea
Therefore, they have common eigenstates and can be diagonalized simultaneously.
Let $|\Phi\rangle$ be a common eigenstate. Acting the transfer matrices on this eigenstate, we have
\bea
&&t(u)|\Psi\rangle=\Lambda(u)|\Psi\rangle,\no\\
&& t^{(1)}(u)|\Psi\rangle= \Lambda^{(1)}(u)|\Psi\rangle,\no\\
&& t^{(2)}(u)|\Psi\rangle= \Lambda^{(2)}(u)|\Psi\rangle.\no
\eea
where $\Lambda(u)$, ${\Lambda}^{(1)}(u)$ and ${\Lambda}^{(2)}(u)$ are the eigenvalues of
$t(u)$, ${t}^{(1)}(u)$ and ${t}^{(2)}(u)$, respectively.
It is easy to check that the eigenvalue $\Lambda(u)$ is a polynomial of $u$ with degree of $2N+2$,
and both ${\Lambda}^{(1)}(u)$ and ${\Lambda}^{(2)}(u)$ are the polynomials of $u$ with degree $2N+4$.
Thus $\Lambda(u)$, ${\Lambda}^{(1)}(u)$ and ${\Lambda}^{(2)}(u)$ can be determined by $6N+13$ independent conditions.

Acting the operator product identities  (\ref{openident1}), (\ref{openident2}) and (\ref{openident3}) on the state $|\Phi\rangle$,
we obtain the functional relations among the eigenvalues
\bea && \Lambda(\pm\theta_j)\Lambda(\pm\theta_j+\eta)=-\frac{1}{
4} \frac{(\pm\theta_j)(\pm\theta_j+\eta)
}{(\pm\theta_j+\frac{1}{{2}}\eta)^2}\nonumber\\[4pt]
&&\hspace{10mm}\times\prod_{l=1}^N
(\pm\theta_j-\theta_l+\eta)(\pm\theta_j+\theta_l+\eta) \Lambda^{(1)}(\pm\theta_j+\frac{1}{2}\eta), \no \\[4pt]
 && \Lambda (\pm\theta_j)\Lambda(\pm\theta_j-\eta)=-\frac{1}{
4} \frac{(\pm\theta_j)(\pm\theta_j-\eta)
}{(\pm\theta_j-\frac{1}{{2}}\eta)^2}\nonumber\\[4pt]
&&\hspace{10mm}\times\prod_{l=1}^N
(\pm\theta_j-\theta_l-\eta)(\pm\theta_j+\theta_l-\eta) \Lambda^{(2)}(\pm\theta_j-\frac{1}{2}\eta), \no \\[4pt]
&& \Lambda (\pm\theta_j-\eta){ \Lambda}^{(1)}(\pm\theta_j+\frac{1}{{2}}\eta)=
\frac{(\pm\theta_j+\frac{1}{2}\eta)^2(\pm\theta_j-\eta)}{(\pm\theta_j+\eta)
(\pm\theta_j-\frac{1}{{2}}\eta)^2}\nonumber\\[4pt]
&&\hspace{10mm}\times
 \prod_{l=1}^N \frac{(\pm\theta_j-\theta_l-\eta)(\pm\theta_j+\theta_l-\eta)}{(\pm\theta_j-\theta_l+\eta)(\pm\theta_j+\theta_l+\eta)} \Lambda (\pm\theta_j+\eta)
 {\Lambda}^{(2)}(\pm\theta_j-\frac{1}{{2}}\eta),\label{eigenident3}
\eea
where $j=1,2,\cdots,N$. Acting Eqs.(\ref{specialvalue4}) and (\ref{openasym3}) on the state $|\Phi\rangle$, we have
\bea
&& \Lambda(0)=0,\quad {\Lambda}^{(1)}(0)=0,\quad {\Lambda}^{(2)}(0)=0, \quad
{\Lambda}^{(1)}(\frac{\eta}{2})=-2\xi \tilde{\xi} \Lambda(\eta), \no \\[4pt]
&&{\Lambda}^{(1)}(-\frac{\eta}{2})=-2\xi \tilde{\xi} \Lambda(-\eta), \quad
{\Lambda}^{(2)}(\frac{\eta}{2})=2\xi \tilde{\xi} \Lambda(\eta),\quad {\Lambda}^{(2)}(-\frac{\eta}{2})=2\xi \tilde{\xi} \Lambda(-\eta),\no\\[4pt]
&& \frac{\partial {\Lambda}^{(1)}(u)}{\partial u}|_{u=0}+ \frac{\partial {\Lambda}^{(2)}(u)}{\partial u}|_{u=0}=0, \no \\[4pt]
&& \Lambda(u)|_{u\rightarrow\infty}=-[c_1\tilde{c}_2+\tilde{c}_1c_2-c_3\tilde{c}_4-\tilde{c}_3c_4] u^{2N+2}, \no \\[4pt]
&& {\Lambda}^{(1)}(u)|_{u\rightarrow\infty}=-4\{2[c_3c_4\tilde{c}_3\tilde{c}_4-\tilde{c_3}c_4-c_3\tilde{c}_4-1]+(1+c_1\tilde{c}_2)^2+(1+\tilde{c_1}c_2)^2\no\\[4pt]
&&\hspace{30mm}-(c_1\tilde{c}_2+\tilde{c}_1c_2)(c_3\tilde{c}_4+\tilde{c}_3c_4)\}u^{2N+4}, \no \\[4pt]
&& {\Lambda}^{(2)}(u)|_{u\rightarrow\infty}=-4\{2[c_1c_2\tilde{c}_1\tilde{c}_2-\tilde{c}_1c_2-c_1\tilde{c}_2-1]+(1+c_3\tilde{c}_4)^2+(1+\tilde{c}_3c_4)^2\no\\[4pt]
&&\hspace{30mm}-(c_1\tilde{c}_2+\tilde{c}_1c_2)(c_3\tilde{c}_4)+\tilde{c}_3c_4\}u^{2N+4}.\label{openasym33}
\eea

Because the operators $\hat U$ given by (\ref{openasym5}) and $\hat Q$ given by (\ref{openasym4}) can be expressed by the constant diagonal matrices, they commutate
with each other and commutate with all the fused transfer matrices. Thus the state $|\Phi\rangle$ also is the eigenvalues of $\hat U$ and $\hat Q$.
After detailed calculation, the operator $\hat U$ has $N+1$ different eigenvalues
\bea
N(2+c_1\tilde{c}_2+\tilde{c}_1c_2)+k(c_3\tilde{c}_4+\tilde{c}_3c_4-c_1
\tilde{c}_2-\tilde{c}_1c_2), \quad k=0,1,\cdots,N.\label{higher-1}
\eea
Eq.(\ref{higher-1}) gives all the possible values of coefficients of the polynomial $\Lambda(u)$ with the degree $2N+1$.
Acting the operator $\hat U$ on the state $|\Phi\rangle$, one would obtain one of them. With direct calculation, we also know the operator
$\hat Q$ has $N+1$ different eigenvalues
\bea
&&N\big[2-2\tilde{c}_1\tilde{c}_2+4c_1\tilde{c}_2+(c_1\tilde{c}_2)^2+4\tilde{c}_1c_2-2c_1c_2+(\tilde{c}_1c_2)^2\big]\nonumber\\[4pt]
&&\hspace{10mm}+k\big[2(c_1\tilde{c}_2+\tilde{c}_1c_2)(c_3\tilde{c}_4+\tilde{c}_3c_4)-2(c_1\tilde{c}_2+\tilde{c}_1c_2)^2\nonumber\\[4pt]
&&\hspace{10mm}+4(c_3\tilde{c}_4+\tilde{c}_3c_4-c_1\tilde{c}_2-\tilde{c}_1c_2)], \quad k=0,1,\cdots N.\label{higher-2}
\eea
Eq.(\ref{higher-2}) indeed gives all the possible values of coefficients of polynomial $\Lambda^{(1)}(u)$ with the degree $2N+3$.
The operator $\hat Q$ acting on the state $|\Phi\rangle$ gives one of them.
Then we arrive at that the above $6N+13$ relations (\ref{eigenident3})-(\ref{higher-2}) enable us to completely determine
the eigenvalues $\Lambda(u)$, ${\Lambda}^{(1)}(u)$ and ${\Lambda}^{(2)}(u)$ which are expressed as the inhomogeneous $T-Q$ relations in the next subsection.

\subsection{Inhomogeneous $T-Q$ relations}

For simplicity, we define $z^{(l)}(u)$, $x_1(u)$ and $x_2(u)$ functions
\begin{eqnarray}
z^{(l)} (u)&=&\left\{
\begin{array}{ll}
\displaystyle(-1)^{p(l)}\alpha_l(u)Q^{(0)}(u)K^{(l)}(u)\frac{Q^{(l-1)}(u+\eta)Q^{(l)}(u-\eta)}{Q^{(l)}(u)Q^{(l-1)}(u)}, &l=1,2,\\[6mm]
\displaystyle(-1)^{p(l)}\alpha_l(u)Q^{(0)}(u)K^{(l)}(u)\frac{Q^{(l-1)}(u-\eta)Q^{(l)}(u+\eta)}{Q^{(l)}(u)Q^{(l-1)}(u)}, &l=3,4,\end{array}
\right. \no \\
x_1 (u)&=&u^2Q^{(0)}(u+\eta)Q^{(0)}(u)\frac{f^{(1)}(u)Q^{(2)}(-u-\eta)}{Q^{(1)}(u)},\no\\[4pt]
x_2 (u)&=&u^2Q^{(0)}(u+\eta)Q^{(0)}(u)Q^{(0)}(-u)\frac{f^{(2)}(u)Q^{(2)}(-u-\eta)}{Q^{(3)}(u)}.\no
\end{eqnarray}
Here the structure factor $\alpha_{l}(u)$ is defined as
\begin{eqnarray}
\alpha_l(u)=\left\{
\begin{array}{ll}
\displaystyle\frac{u}{u+\frac{1}{2}\eta}, &l=1,4,\\[6mm]
\displaystyle\frac{u^2}{(u+\frac{1}{2}\eta)(u+\eta)}, &l=2,3.\end{array}
\right.\no
\end{eqnarray}
The $Q$-functions are
\bea
&&Q^{(0)}(u)=\prod_{l=1}^{N}(u-\theta_l)(u+\theta_l),\quad
Q^{(m)}(u)=\prod_{j=1}^{L_m}(u-\lambda_{j}^{(m)})(u+\lambda_{j}^{(m)}+m\eta), \quad m=1,2,\no \\
&&Q^{(3)}(u)=\prod_{j=1}^{L_3}(u-\lambda_{j}^{(m)})(u+\lambda_{j}^{(m)}+\eta),\quad Q^{(4)}(u)=1,\label{higher-3}
\eea
where $L_1$, $L_2$ and $L_3$ are the non-negative integers which describe the numbers of Bethe roots $\lambda_{j}^{(1)}$, $\lambda_{j}^{(2)}$ and $\lambda_{j}^{(3)}$, respectively.
The forms of functions $K^{(l)}(u)$ are related with the boundary reflections and given by
\bea
&&K^{(1)}(u)=(\xi+\sqrt{1+c_1c_2}u)(\tilde{\xi}+\sqrt{1+\tilde{c}_1\tilde{c}_2}u),\no\\[4pt]
&&K^{(2)}(u)=(\xi-\sqrt{1+c_1c_2}(u+\eta))(\tilde{\xi}-\sqrt{1+\tilde{c}_1\tilde{c}_2}(u+\eta)),\no\\[4pt]
&&K^{(3)}(u)=(\xi+\sqrt{1+c_1c_2}(u+\eta))(\tilde{\xi}+\sqrt{1+\tilde{c}_1\tilde{c}_2}(u+\eta)),\no\\[4pt]
&&K^{(4)}(u)=(\xi-\sqrt{1+c_1c_2}u)(\tilde{\xi}-\sqrt{1+\tilde{c}_1\tilde{c}_2}u).
\eea
The polynomials $f^{(l)}(u)$ in the inhomogeneous terms $x_1(u)$ and $x_2(u)$ are
\bea
f^{(l)}(u)=g_lu(u+\eta)(u-\eta)(u+\frac{1}{2}\eta)^2(u+\frac{3}{2}\eta)(u-\frac{1}{2}\eta)(u+2\eta),\quad l=1,2,\label{func}
\eea
where $g_l$ are given by
\bea
&&g_1=-2-\tilde{c}_1c_2-c_1\tilde{c}_2-2\sqrt{(1+c_1c_2)(1+\tilde{c}_1\tilde{c}_2)},\no\\[4pt]
&&g_2=2+c_3\tilde{c}_4+\tilde{c}_3c_4+2\sqrt{(1+c_1c_2)(1+\tilde{c}_1\tilde{c}_2)}.
\eea

By using the above functions and based on Eqs.(\ref{eigenident3})-(\ref{higher-2}), we construct the eigenvalues $\Lambda(u)$, ${\Lambda}^{(1)}(u)$ and ${\Lambda}^{(2)}(u)$ as following inhomogeneous $T-Q$ relations
\bea
&&\Lambda (u)=\sum_{l=1}^4 z^{(l)} (u)+x_1 (u)+x_2 (u),\no
\\[4pt]
&&\Lambda^{(1)}(u)=-4u^2[Q^{(0)}(u+\frac{1}{2}\eta)(u+\frac{1}{2}\eta)(u-\frac{1}{2}\eta)]^{-1}\Big\{\sum_{l=1}^4\sum_{m=1}^2
\tilde z^{(l)} (u+\frac{1}{2}\eta)\tilde z^{(m)}(u-\frac{1}{2}\eta)\no\\[4pt]
&&\qquad\quad-z^{(1)}(u+\frac{1}{2}\eta)z^{(2)}(u-\frac{1}{2}\eta)+z^{(4)}(u+\frac{1}{2}\eta)z^{(3)}(u-\frac{1}{2}\eta)\Big\}, \no
\\[4pt]
&&\Lambda^{(2)}(u)=-4u^2[Q^{(0)}(u-\frac{1}{2}\eta)(u+\frac{1}{2}\eta)(u-\frac{1}{2}\eta)]^{-1}\Big\{\sum_{l=1}^4\sum_{m=3}^4
\tilde z^{(l)} (u+\frac{1}{2}\eta)\tilde z^{(m)}(u-\frac{1}{2}\eta)\no\\[4pt]
&&\qquad\quad+z^{(1)}(u+\frac{1}{2}\eta)z^{(2)}(u-\frac{1}{2}\eta)-z^{(4)}(u+\frac{1}{2}\eta)z^{(2)}(u-\frac{1}{2}\eta)\Big\},\label{eigen3}
\eea
where
\bea
\tilde z^{(1)}(u)=z^{(1)}(u)+x_1 (u),~\tilde z^{(2)}(u)=z^{(2)}(u),~\tilde z^{(3)}(u)=z^{(3)}(u),~\tilde z^{(4)}(u)=z^{(4)}(u)+x_2 (u).\no
\eea
Since all the eigenvalues are the polynomials, the residues of Eq.(\ref{eigen3}) at the apparent poles should be
zero, which gives the Bethe ansatz equations
\bea &&1+\frac{\lambda_{l}^{(1)}}{\lambda_{l}^{(1)}+\eta}\frac{K^{(2)}(\lambda_{l}^{(1)})Q^{(0)}(\lambda_{l}^{(1)})}{K^{(1)}(\lambda_{l}^{(1)})Q^{(0)}(\lambda_{l}^{(1)}+\eta)}
\frac{Q^{(1)}(\lambda_{l}^{(1)}+\eta)Q^{(2)}(\lambda_{l}^{(1)}-\eta)}{Q^{(1)}(\lambda_{l}^{(1)}-\eta)Q^{(2)}(\lambda_{l}^{(1)})}\no\\[4pt]
&&\qquad=-\frac{\lambda_{l}^{(1)}(\lambda_{l}^{(1)}+\frac{1}{2}\eta)f^{(1)}(\lambda_{l}^{(1)})Q^{(0)}(\lambda_{l}^{(1)})Q^{(2)}(-\lambda_{l}^{(1)}-\eta)}
{K^{(1)}(\lambda_{l}^{(1)})Q^{(1)}(\lambda_{l}^{(1)}-\eta)},\quad l=1,\cdots,L_1,\no\\[4pt]
&&\frac{K^{(3)}(\lambda_{l}^{(2)})}{K^{(2)}(\lambda_{l}^{(2)})}\frac{Q^{(3)}(\lambda_{l}^{(2)}+\eta)}{Q^{(3)}(\lambda_{l}^{(2)})}=
\frac{Q^{(1)}(\lambda_{l}^{(2)}+\eta)}{Q^{(1)}(\lambda_{l}^{(2)})},\quad l=1,\cdots,L_2,\no\\[4pt]
&&\frac{\lambda_{l}^{(3)}(\lambda_{l}^{(3)}+\frac{1}{2}\eta)Q^{(0)}(\lambda_{l}^{(3)}+\eta)Q^{(0)}(-\lambda_{l}^{(3)})f^{(2)}(\lambda_{l}^{(3)})Q^{(2)}(-\lambda_{l}^{(3)}-\eta)}
{K^{(4)}(\lambda_{l}^{(3)})Q^{(3)}(\lambda_{l}^{(3)}-\eta)}\no \\[4pt]
&&\qquad =1+\frac{\lambda_{l}^{(3)}}{\lambda_{l}^{(3)}+\eta}\frac{K^{(3)}(\lambda_{l}^{(3)})}{K^{(4)}(\lambda_{l}^{(3)})}\frac{Q^{(2)}(\lambda_{l}^{(3)}-\eta)Q^{(3)}(\lambda_{l}^{(3)}+\eta)}
{Q^{(2)}(\lambda_{l}^{(3)})Q^{(3)}(\lambda_{l}^{(3)}-\eta)},\quad l=1,\cdots,L_3.\label{open-BAE}
\eea
From the analysis of asymptotic behaviors and contributions of second higher order of corresponding polynomials, the numbers of Bethe roots should satisfy
\bea
L_1=L_2+N+4,\quad  L_3=2N+L_2+4,\quad L_2=k, \quad k=0, 1, \cdots, N.
\eea

Some remarks are in order. The coefficient of term with $u^{2N+1}$ in the polynomial $\Lambda(u)$
and that of term with $u^{2N+3}$ in the polynomial $\Lambda^{(1)}(u)$ are not related with Bethe roots.
The constraints (\ref{higher-1}) and (\ref{higher-2}) require $L_2=k$, where $k=0,\cdots,N$ is related to the eigenvalues of the operators $\hat{U}$ and $\hat{Q}$. Then the Bethe ansatz equations (\ref{open-BAE}) can describe all the eigenstates of the system.
The second set of Bethe ansatz equations in Eq.(\ref{open-BAE}) are the homogeneous ones. This is because that the reflection matrices $K^{(\pm)}(u)$ are the blocking ones.
The matrix elements involving both bosonic (where the parity is 0) and fermionic (where the parity is 1) bases are zero.
The integrability of the system requires that the reflection processes from bosonic basis to fermionic one and vice versa are forbidden.
We note that the Bethe ansatz equations obtained from the regularity of $\Lambda(u)$ are
the same as those obtained from the regularities of $\Lambda^{(1)}(u)$ and $\Lambda^{(2)}(u)$. Meanwhile, the functions $Q^{(m)}(u)$ has two zero points, which
should give the same Bethe ansatz equations.

We have checked that the inhomogeneous $T-Q$ relations (\ref{eigen3}) satisfy the above mentioned $6N+13$ conditions
(\ref{eigenident3})-(\ref{higher-2}). Therefore, $\Lambda(u)$, $\Lambda^{(1)}(u)$ and $\Lambda^{(2)}(u)$ are
the eigenvalues of transfer matrices $t(u)$, ${t}^{(1)}(u)$ and ${t}^{(2)}(u)$, respectively.
Finally, the eigenvalues of Hamiltonian (\ref{hh}) are obtained from $\Lambda(u)$ as
\bea
E=\frac{\partial \ln \Lambda(u)}{\partial u}|_{u=0,\{\theta_j\}=0}.
\eea

\section{Conclusion}

In this paper, we develop a graded nested off-diagonal Bethe ansatz method
and study the exact solutions of the supersymmetric $SU(2|2)$ model with both periodic and off-diagonal boundary conditions.
After generalizing fusion to the supersymmetric case, we obtain the closed sets of operator product identities.
For the periodic case, the eigenvalues are given in terms of the homegeneous $T-Q$ relations (\ref{ep-3}).
While for the open case, the eigenvalues are given by the inhomogeneous $T-Q$ relations (\ref{eigen3}). This scheme can be generalized to other high rank supersymmetric quantum integrable models.

\section*{Acknowledgments}

The financial supports from the National Program for Basic Research of MOST (Grant Nos. 2016YFA0300600 and
2016YFA0302104), the National Natural Science Foundation of China
(Grant Nos. 11934015,
11975183, 11947301, 11774397, 11775178 and 11775177), the Major Basic Research Program of Natural Science of Shaanxi Province
(Grant Nos. 2017KCT-12, 2017ZDJC-32), Australian Research Council (Grant No. DP 190101529), the Strategic Priority Research Program of the Chinese Academy of Sciences (Grant No. XDB33000000), the National Postdoctoral Program for Innovative Talents (BX20180350) and the Double First-Class University Construction Project of Northwest University are gratefully acknowledged.

\section*{Appendix A: Fusion of the reflection matrices}
\setcounter{equation}{0}
\renewcommand{\theequation}{A.\arabic{equation}}

The general fusion procedure of the reflection matrices was given \cite{Mez92, Zho96}. We will generalize the method developed in \cite{Hao14} to study
the fusion of the reflections matrices for super symmetric models (taking the $SU(2|2)$ model as an example). The (graded) reflection equation at special point gives
\begin{equation}
R_{12}(-\alpha){K^{-}_{1}}(u-\alpha)R_{21}(2u-\alpha) {K^{-}_{2}}(u)=
 {K^{-}_{2}}(u)R_{12}(2u-\alpha){K^{-}_{1}}(u-\alpha)R_{21}(-\alpha), \label{oled-1}
 \end{equation}
where $R_{12}(-\alpha)=P_{12}^{(d)}S_{12}$ as we defined perviously. Multiplying Eq.(\ref{oled-1}) with the projector $P_{12}^{(d)}$ from left and using the property $P_{12}^{(d)} R_{12}(-\alpha)= R_{12}(-\alpha)$, we have
\begin{eqnarray}
&&R_{12}(-\alpha){K^{-}_{1}}(u-\alpha)R_{21}(2u-\alpha) {K^{-}_{2}}(u)
\no \\
&&\qquad \qquad =P_{12}^{(d)}
 {K^{-}_{2}}(u)R_{12}(2u-\alpha){K^{-}_{1}}(u-\alpha)R_{21}(-\alpha).\label{oled-2}
\end{eqnarray}
Comparing the right hand sides of Eqs.(\ref{oled-1}) and (\ref{oled-2}), we obtain
\begin{equation}
P_{12}^{(d)} {K^{-}_{2}}(u)R_{12}(2u-\alpha){K^{-}_{1}}(u-\alpha)P_{21}^{(d)}=
 {K^{-}_{2}}(u)R_{12}(2u-\alpha){K^{-}_{1}}(u-\alpha)P_{21}^{(d)}.\label{oled-3}
\end{equation}
Which give the general principle of fusion of the reflection matrices.
If we define $P_{12}^{(d)} {K^{-}_{2}}(u)$ $R_{12}(2u-\alpha){K^{-}_{1}}(u-\alpha)P_{21}^{(d)}$ as the fused reflection matrix $K^-_{\langle 1 2\rangle}(u)\equiv K^-_{\bar 1}(u)$,
where the integrability requires that the inserted $R$-matrix with determined spectral parameter is necessary,
we can prove the the fused $K$-matrix $K^-_{\bar 1}(u)$ also satisfies the (graded) reflection equation
\begin{eqnarray}
&& R_{\bar 12}(u-v) K^{-}_{\bar  1}(u) R _{2\bar 1}(u+v) K^{-}_{2}(v)
=P_{00'}^{(d)}R_{0'2}(u-v)R_{02}(u-v-\alpha)P_{00'}^{(d)}\no \\ &&\qquad
\times P_{00'}^{(d)} {K^{-}_{0'}}(u)R_{00'}(2u-\alpha){K^{-}_{0}}(u-\alpha)P_{0'0}^{(d)}
P_{0'0}^{(d)}R_{20'}(u+v)R_{20}(u+v-\alpha)P_{0'0}^{(d)}  K^{-}_{2}(v)
\no \\ &&
\quad
=P_{00'}^{(d)}R_{0'2}(u-v)R_{02}(u-v-\alpha){K^{-}_{0'}}(u)R_{00'}(2u-\alpha)
\no \\ &&\qquad
\times  {K^{-}_{0}}(u-\alpha)R_{20'}(u+v)R_{20}(u+v-\alpha) K^{-}_{2}(v)P_{0'0}^{(d)}
\no \\ &&\quad
=P_{00'}^{(d)}R_{0'2}(u-v){K^{-}_{0'}}(u)R_{02}(u-v-\alpha)R_{00'}(2u-\alpha)R_{20'}(u+v)
\no \\ &&\qquad
\times  {K^{-}_{0}}(u-\alpha) R_{20}(u+v-\alpha) K^{-}_{2}(v)P_{0'0}^{(d)}
\no \\ &&\quad
=P_{00'}^{(d)}R_{0'2}(u-v){K^{-}_{0'}}(u)R_{20'}(u+v)R_{00'}(2u-\alpha)
\no \\ &&\qquad
\times  R_{02}(u-v-\alpha){K^{-}_{0}}(u-\alpha) R_{20}(u+v-\alpha) K^{-}_{2}(v)P_{0'0}^{(d)}
\no \\ &&\quad
=P_{00'}^{(d)}R_{0'2}(u-v){K^{-}_{0'}}(u)R_{20'}(u+v)R_{00'}(2u-\alpha)
\no \\ &&\qquad
\times K^{-}_{2}(v) R_{02}(u+v-\alpha) {K^{-}_{0}}(u-\alpha) R_{20}(u-v-\alpha) P_{0'0}^{(d)}
\no \\ &&\quad
=P_{00'}^{(d)}R_{0'2}(u-v){K^{-}_{0'}}(u)R_{20'}(u+v)K^{-}_{2}(v)
\no \\ &&\qquad
\times R_{00'}(2u-\alpha) R_{02}(u+v-\alpha) {K^{-}_{0}}(u-\alpha) R_{20}(u-v-\alpha) P_{0'0}^{(d)}
\no \\ &&\quad
=P_{00'}^{(d)}K^{-}_{2}(v) R_{0'2}(u+v) {K^{-}_{0'}}(u) R_{20'}(u-v)
\no \\ &&\qquad
\times R_{00'}(2u-\alpha) R_{02}(u+v-\alpha) {K^{-}_{0}}(u-\alpha) R_{20}(u-v-\alpha) P_{0'0}^{(d)}
\no \\ &&\quad
=K^{-}_{2}(v) P_{00'}^{(d)} R_{0'2}(u+v) {K^{-}_{0'}}(u) R_{02}(u+v-\alpha)R_{00'}(2u-\alpha) R_{20'}(u-v)
\no \\ &&\qquad
\times  {K^{-}_{0}}(u-\alpha) R_{20}(u-v-\alpha) P_{0'0}^{(d)}
\no \\ &&\quad
=K^{-}_{2}(v) P_{00'}^{(d)} R_{0'2}(u+v)  R_{02}(u+v-\alpha){K^{-}_{0'}}(u) R_{00'}(2u-\alpha)  {K^{-}_{0}}(u-\alpha)
\no \\ &&\qquad
\times R_{20'}(u-v) R_{20}(u-v-\alpha) P_{0'0}^{(d)}
\no \\ &&\quad
=K^{-}_{2}(v) R_{\bar 12}(u+v) K^{-}_{\bar 1}(u) R _{2\bar 1}(u-v).
\end{eqnarray}
In the derivation, we have used the relation
\begin{equation}
P_{21}^{(d)}R_{32}(u)R_{31}(u-\alpha)P_{21}^{(d)}=
R_{32}(u)R_{31}(u-\alpha)P_{21}^{(d)}\equiv R_{3\bar 1}(u).\label{so112led-3}
\end{equation}
From the dual reflection equation (\ref{r2}), we obtain the general construction principle of fused dual reflection matrices
\begin{equation}
P_{12}^{(d)} {K^{+}_{2}}(u)R_{12}(-2u-\alpha){K^{+}_{1}}(u+\alpha)P_{21}^{(d)}=
 {K^{+}_{2}}(u)R_{12}(-2u-\alpha){K^{+}_{1}}(u+\alpha)P_{21}^{(d)}.\label{oled-4}
\end{equation}
If $R_{12}(-\beta)=S_{12}P_{12}^{(d)}$, the corresponding fusion relations are
\begin{eqnarray}
&&P_{12}^{(d)} {K^{-}_{1}}(u-\beta)R_{21}(2u-\beta){K^{-}_{2}}(u)P_{21}^{(d)}=
P_{12}^{(d)} {K^{-}_{1}}(u-\beta)R_{21}(2u-\beta){K^{-}_{2}}(u),\label{oled-13}\\[6pt]
&&P_{12}^{(d)} {K^{+}_{1}}(u+\beta)R_{21}(-2u-\beta){K^{+}_{2}}(u)P_{21}^{(d)} \no \\[6pt]
&&\qquad =
P_{12}^{(d)} {K^{+}_{1}}(u+\beta)R_{21}(-2u-\beta){K^{+}_{2}}(u).\label{oled-14}
\end{eqnarray}

Finally the fused K-matrices in subsection 3.2 can be carried out according to Eqs.(\ref{oled-3})-(\ref{oled-4}) or
(\ref{oled-13})-(\ref{oled-14}).

\section*{Appendix B: Proof of the operator product identities}
\setcounter{equation}{0}
\renewcommand{\theequation}{B.\arabic{equation}}
We introduce the reflection monodromy matrices
\bea
&&\hat{T}_{\tilde 0}(u)=R_{N\tilde 0}(u+\theta_N)\cdots R_{2\tilde 0}(u+\theta_2) R_{1\tilde 0}(u+\theta_1) \label{openT5}, \no \\[4pt]
&&\hat{T}_{\tilde 0^\prime}(u)=R_{N\tilde 0^\prime}(u+\theta_N)\cdots R_{2\tilde 0^\prime}(u+\theta_2) R_{1\tilde 0^\prime}(u+\theta_1),\label{openT7}\eea
which satisfy the graded Yang-Baxter equations
\bea
&&R_{1\tilde 2} (u-v) \hat{T}_1(u) \hat{ T}_{\tilde 2}(v) = \hat{ T}_{\tilde 2}(v) \hat{T}_1(u) R_{1\tilde 2} (u-v), \no \\[4pt]
&&R_{1\tilde 2^\prime} (u-v) \hat{T}_1(u) \hat{T}_{\tilde 2^\prime}(v) = \hat{ T}_{\tilde 2^\prime}(v) \hat{T}_1(u) R_{1\tilde 2^\prime} (u-v).\label{yyBb333-1}\eea

In order to solve the transfer matrix $t(u)$ (\ref{tru}), we still need the fused transfer matrices which are defined as
\bea
&&\tilde t^{(1)}(u)= str_{\tilde 0}\{{K}^{+}_{\tilde{0} }(u)  T_{\tilde 0}(u)  {K}^{-}_{\tilde{0} }(u) \hat{T}_{\tilde 0}(u)\}, \no \\[4pt]
&&\tilde t^{(2)}(u)= str_{\tilde 0^\prime}\{{K}^{+}_{\tilde{0}^\prime }(u)   T_{\tilde 0^\prime}(u) {K}^{-}_{\tilde{0} }(u) \hat{ T}_{\tilde 0^\prime}(u)\}.\label{openTransfer-6}
\eea

Similar with periodic case, from the property that above $R$-matrices can degenerate into
the projectors and using the definitions (\ref{openT6}) and (\ref{openT7}),
we obtain following fusion relations among the reflecting monodromy matrices
\bea
&&P^{ (8) }_{21}\hat{T}_2 (u)\hat{T}_1 (u+\eta)P^{(8) }_{21}=\prod_{l=1}^N
(u+\theta_l+\eta)\hat{ T}_{\bar 1}(u+\frac{1}{2}\eta),\no\\[4pt]
&&\bar P^{ (8) }_{21}\hat{T}_2 (u)\hat{T}_1 (u-\eta)\bar P^{(8) }_{21}=\prod_{l=1}^N
(u+\theta_l-\eta)\hat{T}_{\bar 1^\prime }(u-\frac{1}{2}\eta),\no\\[4pt]
&&P^{(20) }_{{\bar 1}2} \hat{T}_{\bar{1}} (u+\frac{1}{2}\eta) \hat{T}_2(u-\eta)P^{(20)
}_{{\bar 1}2}=\prod_{l=1}^N
(u+\theta_l-\eta)\hat{T}_{\tilde 1}(u),\no\\[4pt]
&&P^{(20) }_{{\bar 1}^\prime 2} \hat{T}_{\bar{1}^\prime} (u-\frac{1}{2}\eta)\hat{T}_2(u+\eta)P^{(20)
}_{{\bar 1}^\prime 2}=\prod_{l=1}^N
(u+\theta_l+\eta)\hat{T}_{\tilde 1^\prime }(u).\label{fut-66}\eea

From the definitions, we see that the auxiliary spaces are erased by taking the super partial traces and the physical spaces are the same.
We remark that these transfer matrices are not independent. Substituting Eqs.(\ref{peri-iden}) and (\ref{k-iden}) into the definitions (\ref{openTransfer-6}), we obtain
that the fused transfer matrices $\tilde{t}^{(1)}(u)$ and $\tilde{t}^{(2)}(u)$ are equal
\bea
\tilde{t}^{(1)}(u)=\tilde{t}^{(2)}(u).\label{id0}
\eea
Consider the quantity
\bea
&&t(u)t(u+\eta)
=str_{12}\{K_1^+(u)T_1(u)K_1^-(u)\hat T_1(u)\no\\[4pt]
&&\hspace{12mm}\times[T_2(u+\eta)K_2^-(u+\eta)\hat T_2(u+\eta)]^{st_2}[K_2^+(u+\eta)]^{st_2}\}\no\\[4pt]
&&\hspace{8mm}=[\rho_2(2u+\eta)]^{-1}str_{12}\{K_1^{+}(u)T_1(u)K_1^-(u)\hat T_1(u)\no\\[4pt]
&&\hspace{12mm}\times[T_2(u+\eta)K_2^-(u+\eta)\hat T_2(u+\eta)]^{st_2}R_{21}^{st_2}(2u+\eta)R_{12}^{st_2}(-2u-\eta)[K_2^+(u+\eta)]^{st_2}\}\no\\[4pt]
&&\hspace{8mm}=[\rho_2(2u+\eta)]^{-1}str_{12}\{K_2^+(u+\eta)R_{12}(-2u-\eta)K_1^+(u)T_1(u)T_2(u+\eta)\no\\[4pt]
&&\hspace{12mm}\times K_1^-(u)R_{21}(2u+\eta)K_2^-(u+\eta)\hat T_1(u)\hat T_2(u+\eta)\}\no\\[4pt]
&&\hspace{8mm}=[\rho_2(2u+\eta)]^{-1}str_{12}\{(P_{12}^{(8)}+\bar P_{21}^{(8)})K_2^+(u+\eta)R_{12}(-2u-\eta)K_1^+(u)\no\\[4pt]
&&\hspace{12mm}\times(P_{21}^{(8)}+\bar P_{12}^{(8)}) T_1(u)T_2(u+\eta)(P_{21}^{(8)}+\bar P_{12}^{(8)})K_1^-(u)\no\\[4pt]
&&\hspace{12mm}\times R_{21}(2u+\eta)K_2^-(u+\eta)(P_{12}^{(8)}+\bar P_{21}^{(8)})\hat T_1(u)\hat T_2(u+\eta)(P_{12}^{(8)}+\bar P_{21}^{(8)})\}\no\\[4pt]
&&\hspace{8mm}=[\rho_2(2u+\eta)]^{-1}str_{12}\{[P_{12}^{(8)}K_2^+(u+\eta)R_{12}(-2u-\eta)K_1^+(u)P_{21}^{(8)}]\no\\[4pt]
&&\hspace{12mm}\times [P_{21}^{(8)} T_1(u)T_2(u+\eta)P_{21}^{(8)}]\no\\[4pt]
&&\hspace{12mm}\times [P_{21}^{(8)} K_1^-(u)R_{21}(2u+\eta)K_2^-(u+\eta)P_{12}^{(8)}][P_{12}^{(8)}\hat T_1(u)\hat T_2(u+\eta) P_{12}^{(8)}]\}\no\\[4pt]
&&\hspace{8mm}+[\rho_2(2u+\eta)]^{-1}str_{12}\{[\bar P_{21}^{(8)}K_2^+(u+\eta)R_{12}(-2u-\eta)K_1^+(u)\bar P_{12}^{(8)}]\no\\[4pt]
&&\hspace{12mm}\times [\bar P_{12}^{(8)} T_1(u)T_2(u+\eta)\bar P_{12}^{(8)}]\no\\[4pt]
&&\hspace{12mm}\times[\bar P_{12}^{(8)}K_1^-(u) R_{21}(2u+\eta)K_2^-(u+\eta)\bar P_{21}^{(8)}][\bar P_{21}^{(8)}\hat T_1(u)\hat T_2(u+\eta)\bar P_{21}^{(8)}]\}\no\\[4pt]
&&\hspace{8mm}=t_1(u)+t_2(u).\label{openident1-tan-1}
\eea
The first term is the fusion by the 8-dimensional projectors and the result is
\bea
&&t_1(u)=[\rho_2(2u+\eta)]^{-1}(u+\eta)(u)\prod_{j=1}^{N}(u-\theta_j+\eta)(u+\theta_j+\eta)\no\\[4pt]
&&\hspace{12mm}\times str_{\langle12\rangle}\{K_{\langle12\rangle}^{+}(u+\frac12\eta)T_{\langle12\rangle}^{(8)}(u+\frac12\eta)K_{\langle12\rangle}^{-}(u+\frac12\eta)
\hat T_{\langle12\rangle}^{(8)}(u+\frac12\eta)\}\no\\[4pt]
&&\hspace{8mm}=[\rho_2(2u+\eta)]^{-1}(u+\eta)u\prod_{j=1}^{N}(u-\theta_j+\eta)(u+\theta_j+\eta) t^{(1)}(u+\frac12\eta).
\eea
The second term is the fusion by the other 8-dimensional projectors. Detailed calculation gives
\bea
&&t_2(u)=[\rho_2(2u+\eta)]^{-1}str_{12}\{\bar P_{21}^{(8)}[\bar P_{21}^{(8)}K_2^+(u+\eta)R_{12}(-2u-\eta)K_1^+(u)]\bar P_{12}^{(8)}\no\\[4pt]
&&\hspace{12mm}\times \bar P_{12}^{(8)}[\bar P_{12}^{(8)} T_1(u)T_2(u+\eta)]\bar P_{12}^{(8)}\no\\[4pt]
&&\hspace{12mm}\times\bar P_{12}^{(8)}[\bar P_{12}^{(8)}K_1^-(u) R_{21}(2u+\eta)K_2^-(u+\eta)]\bar P_{21}^{(8)}\no\\[4pt]
&&\hspace{12mm}\times\bar P_{21}^{(8)}[\bar P_{12}^{(8)}\hat T_1(u)\hat T_2(u+\eta)]\bar P_{21}^{(8)}\}\no\\[4pt]
&&\hspace{8mm}=[\rho_2(2u+\eta)]^{-1}str_{12}\{\bar P_{21}^{(8)}[K_1^+(u)R_{21}(-2u-\eta)K_2^+(u+\eta)\bar P_{12}^{(8)}]\bar P_{12}^{(8)}\no\\[4pt]
&&\hspace{12mm}\times \bar P_{12}^{(8)}[T_2(u+\eta)T_1(u)\bar P_{12}^{(8)}]\bar P_{12}^{(8)}\no\\[4pt]
&&\hspace{12mm}\times\bar P_{12}^{(8)}[K_2^-(u+\eta)R_{12}(2u+\eta)K_1^-(u)\bar P_{21}^{(8)}]\bar P_{21}^{(8)}\no\\[4pt]
&&\hspace{12mm}\times\bar P_{21}^{(8)}[\hat T_2(u+\eta)\hat T_1(u)\bar P_{12}^{(8)}]\bar P_{21}^{(8)}\}\no\\[4pt]
&&\hspace{8mm}=[\rho_2(2u+\eta)]^{-1}str_{12}\{[\bar P_{21}^{(8)}K_1^+(u)R_{21}(-2u-\eta)K_2^+(u+\eta)\bar P_{12}^{(8)}]\no\\[4pt]
&&\hspace{12mm}\times [\bar P_{12}^{(8)}T_2(u+\eta)T_1(u)\bar P_{12}^{(8)}]\no\\[4pt]
&&\hspace{12mm}\times[\bar P_{12}^{(8)}K_2^-(u+\eta)R_{12}(2u+\eta)K_1^-(u)\bar P_{21}^{(8)}]\no\\[4pt]
&&\hspace{12mm}\times[\bar P_{21}^{(8)}\hat T_2(u+\eta)\hat T_1(u)\bar P_{21}^{(8)}]\}\no\\[4pt]
&&\hspace{8mm}=[\rho_2(2u+\eta)]^{-1}(u+\eta)u\prod_{j=1}^{N}(u-\theta_j)(u+\theta_j)\no\\[4pt]
&&\hspace{12mm}\times str_{\langle12\rangle^\prime}\{K_{\langle12\rangle^\prime}^{+}(u+\frac12\eta)T_{\langle12\rangle^\prime}(u+\frac 12 \eta)K_{\langle12\rangle^\prime}^{-}(u+\frac12\eta)
\hat T_{\langle12\rangle^\prime}(u+\frac12\eta)\}\no\\[4pt]
&&\hspace{8mm}=[\rho_2(2u+\eta)]^{-1}(u+\eta)u\prod_{j=1}^{N}(u-\theta_j)(u+\theta_j) t^{(2)}(u+\frac12\eta).
\eea
In the derivation, we have used the relations
\bea
&&str_{12}\{A_{12}^{st_1}B_{12}^{st_1}\}=str_{12}\{A_{12}^{st_2}B_{12}^{st_2}\}=str_{12}\{A_{12}B_{12}\},\no\\[4pt]
&&\hat T_1(u)R_{21}(2u+\eta)T_2(u+\eta)=T_2(u+\eta)R_{21}(2u+\eta)\hat T_1(u),\no\\[4pt]
&&P_{12}^{(8)}+\bar P_{12}^{(8)}=1,~~P_{21}^{(8)}+\bar P_{21}^{(8)}=1,~~ P_{12}^{(8)}\bar P_{12}^{(8)}=P_{21}^{(8)}\bar P_{21}^{(8)}=0,~~ P_{12}^{(8)}=P_{21}^{(8)},~~\bar P_{12}^{(8)}=\bar P_{21}^{(8)}.\no
\eea

In addition,
\bea
&& t^{(1)}(u+\frac{1}{2}\eta)t(u-\eta)=str_{\bar 12}\{K_{\bar 1}^{+}(u+\frac{1}{2}\eta)T_{\bar 1}(u+\frac{1}{2}\eta)
K_{\bar 1}^{-}(u+\frac{1}{2}\eta)\hat T_{\bar 1}(u+\frac{1}{2}\eta)\no\\[4pt]
&&\hspace{12mm}\times [T_2(u-\eta)K_2^-(u-\eta)\hat T_2(u-\eta)]^{st_2}[K_2^+(u-\eta)]^{st_2}\}\no\\[4pt]
&&\hspace{8mm}=\rho_4^{-1}(2u-\frac{1}{2}\eta)str_{\bar 12}\{K_{\bar 1}^{+}(u+\frac{1}{2}\eta)T_{\bar 1}
(u+\frac{1}{2}\eta)K_{\bar 1}^{-}(u+\frac{1}{2}\eta)\hat T_{\bar 1}(u+\frac{1}{2}\eta)\no\\[4pt]
&&\hspace{12mm}\times [T_2(u-\eta)K_2^-(u-\eta)\hat T_2(u-\eta)]^{st_2}[R_{2\bar 1}(2u-\frac{1}{2}\eta)]^{st_2}\no\\[4pt]
&&\hspace{12mm}\times[R_{\bar 12}(-2u+\frac{1}{2}\eta)]^{st_2}[K_2^+(u-\eta)]^{st_2}\}\no\\[4pt]
&&\hspace{8mm}=\rho_4^{-1}(2u-\frac{1}{2}\eta)str_{\bar 12}\{K_2^+(u-\eta)R_{\bar 12}(-2u+\frac{1}{2}\eta)K_{\bar 1}^{+}(u+\frac{1}{2}\eta)T_{\bar 1}(u+\frac{1}{2}\eta)\no\\[4pt]
&&\hspace{12mm}\times T_2(u-\eta)K_{\bar 1}^{-}(u+\frac{1}{2}\eta)R_{2\bar 1}(2u-\frac{1}{2}\eta)K_2^{-}(u-\eta)\hat T_{\bar 1}(u+\frac{1}{2}\eta)
\hat T_2(u-\eta)\}\no\\[4pt]
&&\hspace{8mm}=\rho_4^{-1}(2u-\frac{1}{2}\eta)str_{\bar 12}\{(P_{\bar 12}^{(20)}+\tilde P^{(12)}_{\bar 12})K_2^+(u-\eta)R_{\bar 12}(-2u+\frac{1}{2}\eta)
K_{\bar 1}^{+}(u+\frac{1}{2}\eta)\no\\[4pt]
&&\hspace{12mm}\times (P_{2\bar 1}^{(20)}+\tilde P^{(12)}_{2\bar 1})T_{\bar 1}(u+\frac{1}{2}\eta)T_2(u-\eta)(P_{2\bar 1}^{(20)}+\tilde P^{(12)}_{2\bar 1})\no\\[4pt]
&&\hspace{12mm}\times K_{\bar 1}^{-}(u+\frac{1}{2}\eta)R_{2\bar 1}(2u-\frac{1}{2}\eta)K_2^{-}(u-\eta)(P_{\bar 12}^{(20)}+\tilde P^{(12)}_{\bar 12})\no\\[4pt]
&&\hspace{12mm}\times \hat T_{\bar 1}(u+\frac{1}{2}\eta)
\hat T_2(u-\eta)(P_{\bar 12}^{(20)}+\tilde P^{(12)}_{\bar 12})\}\no\\[4pt]
&&\hspace{8mm}=\rho_4^{-1}(2u-\frac{1}{2}\eta)str_{\bar 12}\{P_{\bar 12}^{(20)}K_2^+(u-\eta)R_{\bar 12}(-2u+\frac{1}{2}\eta)K_{\bar 1}^{+}(u+\frac{1}{2}\eta)P_{2\bar 1}^{(20)}\no\\[4pt]
&&\hspace{12mm}\times T_{\bar 1}(u+\frac{1}{2}\eta)T_2(u-\eta)P_{2\bar 1}^{(20)}K_{\bar 1}^{-}(u+\frac{1}{2}\eta)R_{2\bar 1}(2u-\frac{1}{2}\eta)K_2^{-}(u-\eta)\no\\[4pt]
&&\hspace{12mm}\times P_{\bar 12}^{(20)}\hat T_{\bar 1}(u+\frac{1}{2}\eta)
\hat T_2(u-\eta)P_{\bar 12}^{(20)}\}\no\\[4pt]
&&\hspace{12mm}+\rho_4^{-1}(2u-\frac{1}{2}\eta)str_{\bar 12}\{\tilde P_{\bar 12}^{(12)}K_2^+(u-\eta)R_{\bar 12}(-2u+\frac{1}{2}\eta)K_{\bar 1}^{+}(u+\frac{1}{2}\eta)
\tilde P_{2\bar 1}^{(12)}\no\\[4pt]
&&\hspace{12mm}\times T_{\bar 1}(u+\frac{1}{2}\eta)T_2(u-\eta)\tilde P_{2\bar 1}^{(12)}K_{\bar 1}^{-}(u+\frac{1}{2}\eta)R_{2\bar 1}(2u-\frac{1}{2}\eta)K_2^{-}(u-\eta)\no\\[4pt]
&&\hspace{12mm}\times \tilde P_{\bar 12}^{(12)}\hat T_{\bar 1}(u+\frac{1}{2}\eta)
\hat T_2(u-\eta)\tilde P_{\bar 12}^{(12)}\}\no\\[4pt]
&&\hspace{8mm}=\rho_4^{-1}(2u-\frac{1}{2}\eta)(2u+\eta)(u-\eta)\prod_{j=1}^{N}(u-\theta_j-\eta)(u+\theta_j-\eta)\no\\[4pt]
&&\hspace{12mm}\times str_{\langle\bar 12\rangle}\{K_{\langle\bar 12\rangle}^{+}(u)T_{\langle\bar 12\rangle}(u) K_{\langle\bar 12\rangle}^{-}(u)
\hat T_{\langle\bar 12\rangle}(u)\}\no\\[4pt]
&&\hspace{8mm}+\rho_4^{-1}(2u-\frac{1}{2}\eta)(2u+\eta)(u-\eta)\prod_{j=1}^{N}(u-\theta_j)(u+\theta_j)
\no\\[4pt]
&&\hspace{12mm}\times str_{\overline{\langle\bar 12\rangle}}\{ K_{\overline{\langle\bar 12\rangle}}^{+}(u)T_{\overline{\langle\bar 12\rangle}}(u) K_{\overline{\langle\bar 12\rangle}}^{-}(u)
\hat T_{\overline{\langle\bar 12\rangle}}(u)\}\no\\[4pt]
&&\hspace{8mm}=\rho_4^{-1}(2u-\frac{1}{2}\eta)(2u+\eta)(u-\eta)\prod_{j=1}^{N}(u-\theta_j-\eta)(u+\theta_j-\eta)\tilde t^{(1)}(u)\no\\
&&\hspace{12mm}+\rho_4^{-1}(2u-\frac{1}{2}\eta)(2u+\eta)(u-\eta)\prod_{j=1}^{N}(u-\theta_j)(u+\theta_j)\bar{t}^{(1)}(u),\label{tan}\\
&& t^{(2)}(u-\frac{1}{2}\eta)t(u+\eta)=\rho_6^{-1}(2u+\frac{1}{2}\eta)str_{\bar 1^\prime 2}\{K_2^+(u+\eta)R_{\bar 1^\prime 2}(-2u-\frac{1}{2}\eta)\no\\[4pt]
&&\hspace{12mm}\times K_{\bar 1^\prime }^{+}(u-\frac{1}{2}\eta)T_{\bar 1^\prime }(u-\frac{1}{2}\eta) T_2(u+\eta)K_{\bar 1^\prime }^{-}(u-\frac{1}{2}\eta)\no\\[4pt]
&&\hspace{12mm}\times R_{2\bar 1^\prime }(2u+\frac{1}{2}\eta)K_2^{-}(u+\eta)\hat T_{\bar 1^\prime }(u-\frac{1}{2}\eta)
\hat T_2(u+\eta)\}\no\\[4pt]
&&\hspace{8mm}=\rho_6^{-1}(2u-\frac{1}{2}\eta)str_{\bar 1^\prime 2}\{(P_{\bar 1^\prime 2}^{(20)}+\tilde P^{(12)}_{\bar 1^\prime 2})K_2^+(u+\eta)R_{\bar 1^\prime 2}(-2u-\frac{1}{2}\eta)\no\\[4pt]
&&\hspace{12mm}\times K_{\bar 1^\prime }^{+}(u-\frac{1}{2}\eta)(P_{2\bar 1^\prime }^{(20)}+\tilde P^{(12)}_{2\bar 1^\prime })T_{\bar 1^\prime }(u-\frac{1}{2}\eta)
T_2(u+\eta)(P_{2\bar 1^\prime }^{(20)}+\tilde P^{(12)}_{2\bar 1^\prime })\no\\[4pt]
&&\hspace{12mm}\times K_{\bar 1^\prime }^{-}(u-\frac{1}{2}\eta)R_{2\bar 1^\prime }(2u+\frac{1}{2}\eta)K_2^{-}(u+\eta)(P_{\bar 1^\prime 2}^{(20)}
+\tilde P^{(12)}_{\bar 1^\prime 2})\no\\[4pt]
&&\hspace{12mm}\times \hat T_{\bar 1^\prime }(u-\frac{1}{2}\eta)
\hat T_2(u+\eta)(P_{\bar 1^\prime 2}^{(20)}+\tilde P^{(12)}_{\bar 1^\prime 2})\}\no\\[4pt]
&&\hspace{8mm}=\rho_6^{-1}(2u+\frac{1}{2}\eta)(2u-\eta)(u+\eta)\prod_{j=1}^{N}(u-\theta_j+\eta)(u+\theta_j+\eta)\no\\[4pt]
&&\hspace{12mm}\times str_{\langle \bar 1^\prime 2\rangle}\{ K_{\langle\bar 1^\prime 2\rangle}^{+}(u)T_{\langle\bar 1^\prime  2\rangle}(u) K_{\langle\bar 1^\prime 2\rangle}^{-}(u)
\hat T_{\langle\bar 1^\prime 2\rangle}(u)\}\no\\[4pt]
&&\hspace{12mm}+\rho_6^{-1}(2u+\frac{1}{2}\eta)(2u-\eta)(u+\eta)\prod_{j=1}^{N}(u-\theta_j)(u+\theta_j)\no\\[4pt]
&&\hspace{12mm}\times str_{\overline{\langle \bar 1^\prime 2\rangle}}\{K_{\overline{\langle \bar 1^\prime 2\rangle}}^{+}(u)T_{\overline{\langle \bar 1^\prime 2\rangle}}(u)
 K_{\overline{\langle \bar 1^\prime 2\rangle}}^{-}(u)\hat T_{\overline{\langle \bar 1^\prime 2\rangle}}(u)\}\no\\
&&\hspace{8mm}=\rho_6^{-1}(2u+\frac{1}{2}\eta)(2u-\eta)(u+\eta)\prod_{j=1}^{N}(u-\theta_j+\eta)(u+\theta_j+\eta)\tilde t^{(2)}(u)\no\\
&&\hspace{12mm}+\rho_6^{-1}(2u+\frac{1}{2}\eta)(2u-\eta)(u+\eta)\prod_{j=1}^{N}(u-\theta_j)(u+\theta_j)\bar{t}^{(2)}(u),\label{tan-09}
\eea
where we have used the relations
\bea
&&\hat T_{\bar 1}(u+\frac{1}{2}\eta)R_{2\bar 1}(2u-\frac{1}{2}\eta)T_2(u-\eta)=T_2(u-\eta)R_{2\bar 1}(2u-\frac{1}{2}\eta)\hat T_{\bar 1}(u+\frac{1}{2}\eta),\no\\[4pt]
&&P_{\bar 12}^{(20)}+\tilde P_{\bar 12}^{(12)}=1,~~P_{2\bar 1}^{(20)}+\tilde P_{2\bar 1}^{(12)}=1,~~P_{\bar 12}^{(20)}\tilde P_{\bar 12}^{(12)}=0,~~P_{2\bar 1}^{(20)}\tilde P_{2\bar 1}^{(12)}=0, \no \\[4pt]
&&\hat T_{\bar 1^\prime }(u-\frac{1}{2}\eta)R_{2\bar 1^\prime }(2u+\frac{1}{2}\eta)T_2(u+\eta)=T_2(u+\eta)R_{2\bar 1^\prime }(2u+\frac{1}{2}\eta)\hat T_{\bar 1^\prime }(u-\frac{1}{2}\eta),\no\\[4pt]
&&P_{\bar 1^\prime 2}^{(20)}+\tilde P_{\bar 1^\prime 2}^{(20)}=1,~~P_{2\bar 1^\prime }^{(20)}+\tilde P_{2\bar 1^\prime }^{(12)}=1,
~~P_{\bar 1^\prime 2}^{(20)}\tilde P_{\bar 1^\prime 2}^{(12)}=0,~~P_{2\bar 1^\prime }^{(20)}\tilde P_{2\bar 1^\prime }^{(12)}=0.\no
\eea

Focusing on the special points introduced in the main text, we have
\bea
&&t(\pm\theta_j-\eta) t^{(1)}(\pm\theta_j+\frac{1}{2}\eta)=-\frac{1}{2}\frac{(\pm\theta_j+\frac{1}{2}\eta)
(\pm\theta_j-\eta)}{(\pm\theta_j)(\pm\theta_j-\frac{1}{2}\eta)}\no\\
&&\hspace{20mm}\times\prod_{l=1}^{N}(\pm\theta_j-\theta_l-\eta)(\pm\theta_j+\theta_l-\eta)\tilde t^{(1)}(\pm\theta_j),\label{open-ope-1}\\
&&t(\pm\theta_j+\eta) t^{(2)}(\pm\theta_j-\frac{1}{2}\eta)=-\frac{1}{2}\frac{(\pm\theta_j-\frac{1}{2}\eta)
(\pm\theta_j+\eta)}{(\pm\theta_j)(\pm\theta_j+\frac{1}{2}\eta)}\no\\
&&\hspace{20mm}\times\prod_{l=1}^{N}(\pm\theta_j-\theta_l+\eta)(\pm\theta_j+\theta_l+\eta)\tilde t^{(2)}(\pm\theta_j), \quad j=1,2,\cdots,N.\label{open-ope-2}
\eea
With the help of  Eqs. (\ref{id0}), (\ref{open-ope-1}) and (\ref{open-ope-2}), we can derive the relation (\ref{openident3}).
Finally, we have proven the identities (\ref{openident1})-(\ref{openident3}).


\end{document}